\documentclass[cernpreprint, texlive=2020, UKenglish, texmf]{atlasdoc}

\usepackage[subcaption]{atlaspackage}
\usepackage{atlasbiblatex}
 
\usepackage{atlasphysics}
 
\graphicspath{{logos/}{figures/}}   
 
\usepackage{ANA-EXOT-2018-29-PAPER-defs}

\AtlasTitle{A search for an unexpected asymmetry in the production of $e^+ \mu^-$ and $e^- \mu^+$ pairs in proton--proton collisions recorded by the ATLAS detector at $\sqrt s = 13$\,TeV}
 
\AtlasAbstract{
This search, a type not previously performed at ATLAS, uses a comparison of the production cross sections for $e^+ \mu^-$ and $e^- \mu^+$ pairs to constrain physics processes beyond the Standard Model.
It uses $139\,\text{fb}^{-1}$ of proton--proton collision data recorded at $\sqrt{s} = 13$\,TeV at the LHC.
Targeting sources of new physics which prefer final states containing $e^{+}\mu^{-}$ to $e^{-}\mu^{+}$, the search contains two broad signal regions which are used to provide model-independent constraints on the ratio of cross sections at the 2\% level.
The search also has two special selections targeting supersymmetric models and leptoquark signatures.
Observations using one of these selections are able to exclude, at 95\% confidence level, singly produced smuons with masses up to 640\,GeV in a model in which the only other light sparticle is a neutralino when the $R$-parity-violating coupling $\lambda'_{231}$ is close to unity.
Observations using the other selection exclude scalar leptoquarks with masses below 1880\,GeV when $g_{\text{1R}}^{eu}=g_{\text{1R}}^{\mu c}=1$, at 95\% confidence level.
The limit on the coupling reduces to $g_{\text{1R}}^{eu}=g_{\text{1R}}^{\mu c}=0.46$ for a mass of 1420\,GeV.
}
 
\author{The ATLAS Collaboration}
\AtlasJournalRef{Phys. Lett. B 830 (2022) 137106}
\AtlasDOI{10.1016/j.physletb.2022.137106}
\PreprintIdNumber{CERN-EP-2021-190}
 
\hypersetup{pdftitle={ATLAS document},pdfauthor={The ATLAS Collaboration}}
 
\begin{document}

\newcommand\draftIsInOneColumnFormat{}

\maketitle

\ifOneOrTwoCol{}{\twocolumn}

\section{Introduction}
Under the Standard Model of particle physics, the ratio of the production cross section of a positron and muon together in a proton--proton interaction is expected to be very similar to the production cross section for an electron and anti-muon.
This similarity is a consequence of the lepton flavour universality of the electroweak boson interactions that produce these leptons, in combination with charge conservation and the relatively low mass of these leptons with respect to the mass of the electroweak bosons.
As was explored in Ref.~\cite{Lester:2016qdv}, measuring the ratio of these cross sections can serve as an experimental test of this aspect of the Standard Model (SM) and could have sensitivity to physics Beyond the Standard Model (BSM).
Specifically, the ratio:
\ifOneOrTwoCol{\vspace{-1mm}}{}
\begin{equation}
\ourRatio \equiv  \frac{\sigma(pp\rightarrow e^+\mu^-+X)}{\sigma(pp\rightarrow e^-\mu^++X)}\label{eq:defr}
\end{equation}
is defined, where the leptons are all taken to be promptly produced in the primary interaction\footnote{The \(X\) in Eq.~\eqref{eq:defr} contains no further prompt leptons.}. Scenarios are considered where the presence of a BSM process would bias \(\ourRatio\) to be significantly greater than one (more \emum\ than \emup).
Two concrete examples of such BSM models are considered in this \paperspecies. The first is an \(R\)-parity-violating supersymmetry model.
As was noted in Ref.~\cite{Lester:2016qdv}, a non-zero \(R\)-parity-violating \(\lptwothreeone\) coupling (defined in Refs.~\cite{PhysRevD.40.2987,CHOUDHURY1996201,Dreiner:1997uz}) linking smuons to top and down quarks could easily drive
\(\ourRatio>1\) as the proton's down quarks would result in
\ifOneOrTwoCol{\vspace{-4mm}}{\vspace{-2mm}}
\begin{equation}\nonumber\hbox{
\begin{tikzpicture}[baseline=(current  bounding  box.center),
scale=1,
thick,
level/.style={level distance=3.15cm, line width=0.4mm},
level 2/.style={sibling angle=60},
level 3/.style={sibling angle=60},
level 4/.style={level distance=1.4cm, sibling angle=60}
]
\draw[jblue,very thick,electron] (-1.2,0.5) -- (0,0.5) ;
\draw[jblue,very thick,electron] (0,0.5) -- (0,-1) ;
\draw[jblue,very thick,electron] (0,-1) -- (1,-1) ;
\draw[jblue,very thick,electron] (2,-0.5) -- (1,-1) ;
\draw[jblue,very thick,electron] (1,-1) -- (2,-1.0) ;
\draw[jblue,very thick,electron] (1,-1) -- (2,-1.5) ;
\draw[jblue,very thick,dashed] (0,0.5) -- (1,0.5) ;
\draw[jblue,very thick,electron] (1,0.5) -- (2,+1) ;
\draw[jblue,very thick,photon] (1,0.5) -- (2,0) ;
\draw[jblue,very thick,straightBit] (1,0.5) -- (2,0) ;
\draw[jblue,very thick,gluon] (-1.2,-1) -- (0,-1) ;
\node[draw=none,fill=none] at (-1.4,0.5){\(d\)}  ;
\node[draw=none,fill=none] at (0.5,+0.9){\(\tilde \mu_\text{L}^-\)}  ;
\node[draw=none,fill=none] at (-0.3,-0.25){\(t\)}  ;
\node[draw=none,fill=none] at (-1.4,-1){\(g\)}  ;
\node[draw=none,fill=none] at (2.3,+1.1){\(\mu^-\)}  ;
\node[draw=none,fill=none] at (2.3,-0){\(\chi^0_1\)}  ;
\node[draw=none,fill=none] at (0.5,-0.60){\(t\)}  ;
\node[draw=none,fill=none] at (2.3,-0.5){\(e^+\)}  ;
\node[draw=none,fill=none] at (2.3,-1){\(\nu_e\)}  ;
\node[draw=none,fill=none] at (2.3,-1.5){\(b\)}  ;
\end{tikzpicture}
}\end{equation} 
\ifOneOrTwoCol{}{}
(and related s-channel processes) occurring more frequently than the charge conjugate process:
\ifOneOrTwoCol{}{\vspace{-2mm}}
\begin{equation}\nonumber\hbox{
\begin{tikzpicture}[baseline=(current  bounding  box.center),
scale=1,
thick,
level/.style={level distance=3.15cm, line width=0.4mm},
level 2/.style={sibling angle=60},
level 3/.style={sibling angle=60},
level 4/.style={level distance=1.4cm, sibling angle=60}
]
\draw[jblue,very thick,electron] (0,0.5) -- (-1.2,0.5) ;
\draw[jblue,very thick,electron] (0,-1) -- (0,0.5)  ;
\draw[jblue,very thick,electron] (1,-1) -- (0,-1);
\draw[jblue,very thick,electron] (1,-1) -- (2,-0.5) ;
\draw[jblue,very thick,electron] (2,-1.0) -- (1,-1) ;
\draw[jblue,very thick,electron] (2,-1.5) -- (1,-1);
\draw[jblue,very thick,dashed] (0,0.5) -- (1,0.5) ;
\draw[jblue,very thick,electron] (2,+1) -- (1,0.5) ;
\draw[jblue,very thick,straightBit2] (1,0.5) -- (2,0) ;
\draw[jblue,very thick,photon] (1,0.5) -- (2,0) ;
\draw[jblue,very thick,gluon] (-1.2,-1) -- (0,-1) ;
\node[draw=none,fill=none] at (-1.4,0.5){\(\bar d\)}  ;
\node[draw=none,fill=none] at (0.5,+0.9){\(\tilde \mu_\text{L}^+\)}  ;
\node[draw=none,fill=none] at (-0.3,-0.25){\(t\)}  ;
\node[draw=none,fill=none] at (-1.4,-1){\(g\)}  ;
\node[draw=none,fill=none] at (2.3,+1.1){\(\mu^+\)}  ;
\node[draw=none,fill=none] at (2.3,-0){\(\chi^0_1\)}  ;
\node[draw=none,fill=none] at (0.5,-0.60){\(\bar t\)}  ;
\node[draw=none,fill=none] at (2.3,-0.5){\(e^-\)}  ;
\node[draw=none,fill=none] at (2.3,-1){\(\bar\nu_e\)}  ;
\node[draw=none,fill=none] at (2.3,-1.5){\(\bar b\).}  ; 
\end{tikzpicture}
}
\end{equation}
\ifOneOrTwoCol{}{}
The second model considered to drive \(\ourRatio>1\) is a scalar leptoquark with couplings permitting \(S_1\rightarrow ue^-\) and \(S_1\rightarrow c\mu^-\).
In that case, processes such as
\ifOneOrTwoCol{\vspace{-3mm}}{\vspace{-5mm}}
\begin{equation}\nonumber\hbox{
\begin{tikzpicture}[baseline=(current  bounding  box.center), scale=1]
\draw[jblue,very thick,electron] (-1.2,0.5) -- (-0.5,-0.25) ;
\draw[jblue,very thick,gluon   ] (-1.2,-1.) -- (-0.5,-0.25) ;
\draw[jblue,very thick,electron] (-0.5,-0.25) -- (0.25,-0.25) ;
\draw[jblue,very thick,dashed] (0.25,-0.25) -- (1,0.5) ;
\draw[jblue,very thick,electron] (1,0.5) -- (2,+1) ;
\draw[jblue,very thick,electron] (1,0.5) -- (2,0) ;
\draw[jblue,very thick,electron] (2,-1) -- (0.25,-0.25) ;
 
\node[draw=none,fill=none] at (-1.4,0.5){\(u\)}  ;
\node[draw=none,fill=none] at (0.4,0.4){\(S_1\)}  ;
\node[draw=none,fill=none] at (0,-0.5){\(u\)}  ;
\node[draw=none,fill=none] at (-1.4,-1){\(g\)}  ;
\node[draw=none,fill=none] at (2.3,+1.1){\(\mu^-\)}  ;
\node[draw=none,fill=none] at (2.3,-0){\(c\)}  ;
\node[draw=none,fill=none] at (2.3,-1){\(e^+\)}  ;
\end{tikzpicture}
}\end{equation}
would be favoured over charge conjugates.
 
This \paperspecies\ describes a measurement of \(\ourRatio\) using the ATLAS detector at the LHC, which is used as a test for the existence of BSM physics that would bias \(\ourRatio\) above one.
While related BSM processes could also bias \(\ourRatio\) to be significantly less than one, sensitivity to such processes are not considered in this \paperspecies\ because of the existence of experimental effects that predominantly bias the measured ratio downwards.
Examples of these effects are explored further in Ref.~\cite{Lester:2016qdv}, but one such example is the presence of hadronic jets being incorrectly reconstructed as an electron more frequently than as a muon.
This bias for fake electrons over fake muons, in combination with the predominance of \(W^{+}\) over \(W^{-}\) bosons from the \((pp)^{2+}\) initial state~\cite{CMS-SMP-18-012,STDM-2015-03}, would manifest as a bias for \emup\ (with a fake \(e^{-}\)) over \emum\ (with a fake \(\mu^{-}\)), and hence bias the measured \(\ourRatio\) downwards.
The analysis strategy presented in this \paperspecies\ includes estimating corrections to account for \emph{some} of these biasing experimental effects, the primary motivation for which is the enhancement of sensitivity to BSM physics that increases \(\ourRatio\) rather than to improve the accuracy of the \(\ourRatio\) measurement, and estimating systematic uncertainties to account for any remaining biases.
In particular, for the model-independent statistical analysis of the measured \(\ourRatio\) in this \paperspecies\ it is assumed that the SM hypothesis covers values of \(\ourRatio\) less than or equal to one, an assumption that is checked in data in the analysis event selections\footnote{See definitions of \crratio\ and \crjet\ in Section~\ref{sec:analysis}, and the estimates of fake-lepton contributions to Standard Model expectations.}.
Thus evidence for new physics would only be claimed if the BSM contribution is strong enough to overcome any residual biases that have not been accounted for. For calculating limits on signal model parameters the strength of the residual biases are estimated from an SM-enriched control region.
 
The structure of this \paperspecies\ is as follows: in Section~\ref{sec:detector} the ATLAS detector is described and Section~\ref{sec:data} details the datasets used in the analysis. Section~\ref{sec:objects} describes the object reconstruction and Section~\ref{sec:analysis} defines the regions of phase space that \(\ourRatio\) is measured in and how experimental effects that could impact its measurement are corrected for, including the verification of these corrections with the measurement of \(\ourRatio\) in SM-enriched control regions.
Section~\ref{sec:result} presents the measurement of \(\ourRatio\) in generic signal regions designed to have broad sensitivity to BSM physics that would bias \(\ourRatio\) upwards, then in optimised signal regions designed to have sensitivity to the two BSM models described above, alongside a statistical interpretation of the result in the parameter space of these models.
\section{ATLAS detector}
\label{sec:detector}
\newcommand{\AtlasCoordFootnote}{
ATLAS uses a right-handed coordinate system with its origin at the nominal interaction point (IP)
in the centre of the detector and the \(z\)-axis along the beam pipe.
The \(x\)-axis points from the IP to the centre of the LHC ring, and the \(y\)-axis points upwards.
Cylindrical coordinates \((r,\phi)\) are used in the transverse plane, \(\phi\) being the azimuthal angle around the \(z\)-axis.
The pseudorapidity is defined in terms of the polar angle \(\theta\) as \(\eta = -\ln \tan(\theta/2)\).
Angular distance is measured in units of \(\Delta R \equiv \sqrt{(\Delta\eta)^{2} + (\Delta\phi)^{2}}\).}
 
The ATLAS experiment~\cite{PERF-2007-01} at the LHC is a multipurpose particle detector with a forward--backward symmetric cylindrical geometry and a near \(4\pi\) coverage in solid angle.\footnote{\AtlasCoordFootnote}
It consists of an inner tracking detector surrounded by a thin superconducting solenoid providing a \SI{2}{\tesla} axial magnetic field, electromagnetic and hadron calorimeters, and a muon spectrometer.
The inner tracking detector covers the pseudorapidity range \(|\eta| < 2.5\).
It consists of silicon pixel, silicon microstrip, and transition radiation tracking detectors.
Lead/liquid-argon (LAr) sampling calorimeters provide electromagnetic (EM) energy measurements with high granularity.
A steel/scintillator-tile hadron calorimeter covers the central pseudorapidity range (\(|\eta| < 1.7\)).
The endcap and forward regions are instrumented with LAr calorimeters for both the EM and hadronic energy measurements up to \(|\eta| = 4.9\).
The muon spectrometer surrounds the calorimeters and is based on three large superconducting air-core toroidal magnets with eight coils each.
The field integral of the toroids ranges between \num{2.0} and \SI{6.0}{\tesla\(\cdot\)\metre} across most of the detector.
The muon spectrometer includes a system of precision chambers for tracking and fast detectors for triggering.
A two-level trigger system is used to select events.
The first-level trigger is implemented in hardware and uses a subset of the detector information to accept events at a rate below \SI{100}{\kHz}.
This is followed by a software-based trigger that reduces the accepted event rate to \SI{1}{\kHz} on average depending on the data-taking conditions.
 
An extensive software suite~\cite{ATL-SOFT-PUB-2021-001} is used in the reconstruction and analysis of real and simulated data, in detector operations, and in the trigger and data acquisition systems of the experiment.
 
\section{Data and Monte Carlo samples}
\label{sec:data}
The proton--proton collisions analysed in this \paperspecies\ are those collected at a centre-of-mass energy of \(\sqrt s=13~\TeV\) and a 25~ns interbunch spacing between 2015 and 2018.
They correspond to an integrated luminosity of \(\INTLUMIIFB~\ifb\).
The uncertainty in the combined 2015--2018 integrated luminosity is 1.7 \% \cite{ATLAS-CONF-2019-021}, obtained using the LUCID-2 detector \cite{LUCID2} for the primary luminosity measurements.
In any given data-taking period the unprescaled two-lepton triggers (specifically \(ee\), \(e\mu\) or \(\mu\mu\)) with the lowest per-lepton \(p_\text{T}\) thresholds were used~\cite{TRIG-2018-05,TRIG-2018-01,ATLAS:2021lws}. These thresholds ranged from 10~\GeV\ to 24~\GeV.
 
\(R\)-parity-violating (RPV) \cite{Farrar:1978xj} models of supersymmetry \cite{Golfand:1971iw,Volkov:1973ix,Wess:1974tw,Wess:1974jb,Ferrara:1974pu,Salam:1974ig} were tested using simulated events with \(\mu^{-}\tilde{\chi}^{0}_{1} t\) or \(\mu^{+}\tilde{\chi}^{0}_{1} \bar{t}\) in the final state, where \(\tilde{\chi}_{1}^{0}\) is the lightest neutralino.
This neutralino is considered stable enough on detector scales that it can only be detected through missing transverse momentum, unless it approaches or exceeds the top-quark mass, when it can decay through the RPV coupling.
These events were generated at leading order using the Monte Carlo (MC) program \MGFiveAMCatNLO\ \CiteRef{Madgraph}\ version 2.61 together with the \texttt{RPV MSSM UFO} model~\cite{rpvmssmUfo}.
Shower evolution and hadronisation was performed by \PythiaEight~\CiteRef{PythiaEight}\ version 8.23. The \NNPDFTwoThree~\cite{Ball:2012cx} PDF was used with a set of tuned parameters called the A14 tune \cite{ATL-PHYS-PUB-2014-021}.
All RPV couplings except \lptwothreeone\ were set to zero, so only the smuon RPV interaction is considered.
Supersymmetric particles other than the neutralino and smuon were decoupled by setting their masses to a large value.
The \textsc{Madgraph} hard processes permitted at most one additional light parton in the final state, and they were
matched to the \Pythia\ parton shower using the CKKW-L~\cite{Lonnblad:2012ix} merging scheme.
Use of the matching scale \(Q_\text{MS}=\frac{1}{4} \big(m(t) + m(\tilde\chi^0_1)\big)\)
gives a smooth transition between the matrix-element and parton-shower regimes, and distributions with little dependence on the exact scale value.
Event samples were generated for a two-dimensional grid of points, distributed in a plane of smuon and neutralino masses, all with a coupling of \(\lptwothreeone=1.0\).  Samples for other values of the coupling were obtained by weighting the cross sections of the first set in proportion to the square of the desired value of \lptwothreeone\ and changing the branching ratio for the smuon decay.
The branching ratio for the desired smuon to muon decay is 2\% (70\%) at \(\lptwothreeone=1\) (0.1), whilst the remaining smuons decay via the RPV vertex.
 
Leptoquark events were generated at leading order using \MGFiveAMCatNLO\,2.61 together with the `\(S_1\)' model of Ref.~\cite{BUCHMULLER1987442},
which is implemented as a Feynrules \cite{Alloul:2013bka} package named `\texttt{LO\_LQ\_S1}' available
online~\cite{L0LQS1package} and described in Ref.~\cite{Dorner2018}.
Shower evolution and hadronisation were performed by \Pythia\,8.23. The \NNPDFTwoThree\ PDF was used with the A14 tune.
All leptoquark couplings were set to zero, apart from two flavours of the \(g_\textrm{1R}\) coupling of Ref.~\cite{BUCHMULLER1987442} which couples leptoquarks to leptons and quarks in weak singlet states.  Specifically, \(g_\textrm{1R}^{eu}\) and \(g_\textrm{1R}^{\mu c}\) were assigned a common non-zero value simply denoted `\(\lambda\)'.
The leptoquark signal events were generated for a set of leptoquark masses, all with a coupling of \(\lambda=1.0\).
Charm-quark-initiated processes are neglected since they provide no charge--flavour asymmetry, and their cross section is only \(\mathcal{O}(5\%)\) of that for up-quark-initiated processes.
The hard processes specified no additional light jets in the final state and they were matched to the \Pythia\ parton shower using the CKKW-L~\cite{Lonnblad:2001iq} merging scheme.
The merging scale \(Q_\text{MS}\) was chosen to be \(\frac{1}{4} \big(m(t) + m(S_1)\big)\), where \(m(S_1)\) is the mass of the leptoquark.
 
Next-to-leading order (NLO) cross sections are used for the samples and were calculated using \MGFiveAMCatNLO\,2.94 with a
narrow-width (nw) approximation and the \NNPDFTwoThreeNLO\ PDF set.
To include the effect of the finite-width (fw) approximation in the samples, the cross sections are corrected thus:
\begin{equation}
\sigma_\text{Total}^\text{NLO,fw} = \frac{\sigma_{\emum}^\text{NLO,nw}}{\sigma_{\emum}^\text{LO,nw}}\sigma_{\emum}^\text{LO,fw} + \frac{\sigma_{\emup}^\text{NLO,nw}}{\sigma_{\emup}^\text{LO,nw}}\sigma_{\emup}^\text{LO,fw},
\end{equation}
following an approach similar to that used in Ref.~\cite{Roy:2020fqf}.
The narrow-width leading order (LO) cross sections are calculated using \MGFiveAMCatNLO\,2.94 and the \NNPDFTwoThree\ PDF set.
The finite-width LO cross sections are calculated using the same set-up as the generated samples described above.
Theoretical uncertainties in the NLO cross sections are calculated by \MGFiveAMCatNLO\ using an envelope of nine variations of the renormalisation and factorisation scales, each taking values of either 0.5, 1,0 or 2.0 times their nominal value.
 
Samples for other values of the coupling \(\lambda\) were obtained by weighting the cross sections of the first set in proportion to the square of the desired value of \(\lambda\) and accounting for the change in leptoquark width.
A two-dimensional grid of samples for a variety of leptoquark masses and couplings was thereby obtained.
All signal sample events were then simulated using \AtlFast~\cite{ATLAS:2010ytq}, a fast simulation of the ATLAS detector.
 
MC simulations of SM processes are not used in the final result of this analysis, but were used to guide the signal-region choices, to study the validity of the analysis strategy,
to assist in deriving efficiencies and uncertainties for the fake-lepton background estimate,
and to perform cross-checks (see Appendix~\appsmmc\ for MC sample details). Instead, measurements of \(\ourRatio\) are based entirely on comparisons between \(e^+\mu^-\) and \(e^-\mu^+\) data, and contributions from jets misidentified as leptons, muon corrections and even expected SM yields (see Section~\ref{sec:result}) are also estimated primarily from data.
\section{Reconstructed objects}
\label{sec:objects}
Reconstructed objects (electrons, muons, jets, missing transverse momentum) are the building blocks of any analysis.
In this analysis, leptons and jets exist in two forms: `\dull' and `\polished'.
The former are used to define missing transverse momentum and the procedure to resolve ambiguities between objects with overlapping constituents, otherwise the analysis regions are built exclusively on the latter.
 
\dull\ electrons are required to have \(|\eta|<2.47\) and \(\pt>10~\GeV\), and to pass the \texttt{Loose} likelihood-based identification working point defined in Ref.~\cite{EGAM-2018-01}.
The same \pt\ and \(|\eta|\) demands are placed on \dull\ muons, which are also required to pass the \texttt{Medium} identification working point as defined in Ref~\cite{MuonPerfFullRun2}.
The \antikt\ algorithm \cite{Cacciari:2008gp,Fastjet} with a radius parameter of \(R=0.4\) is used to reconstruct jets with a four-momentum recombination scheme, using `particle-flow' objects \cite{Aaboud:2017aca} as inputs.
\dull\ jets are required to have \(\pt>20~\)\GeV\ and \(|\eta|<4.5\).
The missing transverse momentum, \(\vecptmiss\), is calculated from the \dull\ leptons and jets as described in Ref.~\cite{ATLAS-CONF-2018-023} using the \texttt{Tight} working point and `particle-flow track-based soft term' defined therein.
 
An overlap removal procedure is applied to \dull\ jets and \dull\ leptons to avoid double-counting.
Firstly, any electron which shares a track with a muon is rejected.
Any jet whose angular distance $\Delta R$ from an electron is less than 0.2
is removed, as is any which has fewer than three tracks lying within \(\Delta R = 0.4\) of a muon.
Finally, electrons and muons within \(\Delta R = 0.4\) of the remaining jets are then discarded.
 
The `jet vertex tagger' (JVT)~\cite{ATLAS-CONF-2014-018} is applied to jets with \(|\eta|<2.4\) and \(\pt<120~\GeV\), and helps to veto jets that are likely to have originated from pile-up (additional \(pp\) collisions from the same or nearby bunch crossings).
A similar `forward jet vertex tagger' (fJVT) is used to help identify and remove pile-up jets with \(|\eta|>2.5\)~\cite{fixedPERF-2016-06}.
Jets surviving the overlap removal procedure are deemed \polished\ if they pass the JVT or fJVT, and have \(|\eta|<2.8\).
 
Those leptons which remain are then given a status of \polished\ if they meet the following five criteria:
(i) they must have \(\pt>25~\GeV\) and \(|\eta|<2.47\), and
(ii) be consistent with the hard-scatter vertex, through \(|d_0/\sigma(d_0)|<3\) and \(|z_0\sin(\theta)|<0.3\)\,mm, where \(d_0\) and \(z_0\) are their transverse and longitudinal impact parameters;
(iii) electrons must pass the \texttt{Tight} likelihood-based identification working point defined in Ref.~\cite{EGAM-2018-01}, and have charge misidentification suppressed through the use of the boosted-decision-tree-based discriminant described in Ref.~\cite{EGAM-2018-01};
(iv) electrons with \(\pt<200~\GeV\) (\(\pt>200~\GeV\)) are required to pass the \texttt{Tight} (\texttt{HighPtCaloOnly}) isolation working point of Ref.~\cite{EGAM-2018-01} to reduce contamination by electrons from heavy-flavour decays or misidentified light hadrons; and
(v) muons with \(\pt<200~\GeV\) (\(\pt>200~\GeV\)) are required to pass the \texttt{Tight} (\texttt{FixedCutHighPtTrackOnly}) isolation working point of Ref.~\cite{MuonPerfFullRun2} to reduce contamination by muons from semileptonic heavy-flavour and hadron decays.
 
In the rest of this \paperspecies, leptons and jets are assumed to be only those with \polished\ status, unless stated otherwise.
\section{Analysis}
\label{sec:analysis}
The tests of whether \(\ourRatio\) is significantly greater than one are made in four signal regions.
Two of them, (\srmet\ and \srjet) aim to provide sensitivity to general sources of charge--flavour violation, while the other two (\srrpv\ and \srlq) are less inclusive versions of their partners\footnote{Every event in \srrpv\ is also in \srmet, and every event in \srlq\ is also in \srjet.} and target specific RPV supersymmetry and leptoquark theories mentioned in the introduction.
These regions are summarised in Figure~\ref{fig:regions}.
 
\begin{figure}
\centering
\includegraphics[width=\ifOneOrTwoCol{0.6\linewidth}{0.99\linewidth}]{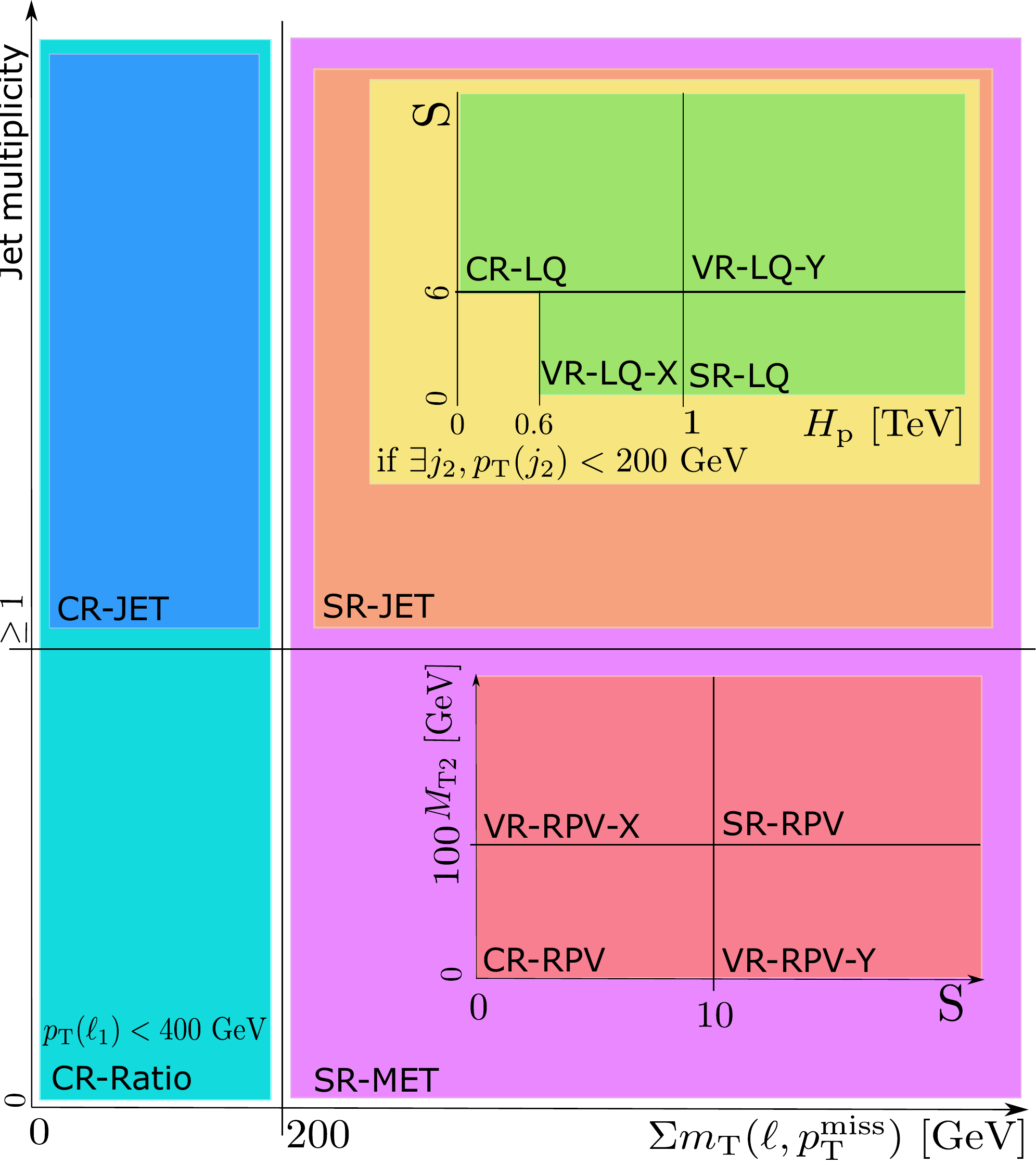}
\caption{Summary of the regions used in the analysis.
\(\ell_1\) refers to the lepton with the largest \pt\ in each event.
\(j_2\) refers to the jet with the second-largest \pt\ in each event.
\srmet\ and \srjet\ are used for the \(\ourRatio\) measurement.
\crratio\ and \crjet\ are used to check the SM expectation of \(\ourRatio\leq1\), and \crratio\ is used to derive the residual-bias uncertainty included in the ratio measurement.
\crrpv\ is used to derive the SM background expectation used to set limits on RPV supersymmetry models in \srrpv, after being validated in \vrrpvx\ and \vrrpvy.
\crlq\ is used to derive the SM background expectation used to set limits on scalar leptoquark models in \srlq, after being validated in \vrlqx\ and \vrlqy.
Note that the horizontal positions of the boxes do not denote \SUMMT\ thresholds beyond the \(200~\GeV\) cut.}
\label{fig:regions}
\end{figure}
 
The \baselineSELECTIONNOTLEPTON\ selection common to all signal regions requires the presence of exactly one electron and exactly one muon, of opposite charge.
As there are few other constraints on the forms that the signal regions should take, and as the ATLAS experiment has not previously published a charge--flavour asymmetry search, the approach taken in this first search is to prioritise simple selections over complex ones.
With this principle in mind:
\begin{itemize}
\item
when defining \srmet, the only requirement which is added to the \baselineSELECTIONNOTLEPTON\ selection is that \(\SUMMT>200~\GeV\), where \(\SUMMT\equiv m_\mathrm{T}(e,\vecptmiss)+m_\mathrm{T}(\mu,\vecptmiss)\) and \(m_\mathrm{T}\) is the usual transverse mass: \[m_\mathrm{T}(\ell,\vecptmiss)\equiv \sqrt{2 |\vec p^{\ell}_\mathrm{T}|     |\vecptmiss|  - 2 \vec p^{\ell}_\mathrm{T} \cdot
\vecptmiss
},\]
\item
and when defining \srjet\ (a subset of \srmet) the only additional requirement is that events must contain at least one jet with transverse momentum greater than \(20~\GeV\).
\end{itemize}
The primary purpose of the \SUMMT\ requirement is to allow for the low \SUMMT\ region to be available for studying the SM behaviour before unblinding the data in the signal regions.
The benchmark BSM models have a low yield in \(\SUMMT<200~\)\GeV, and it can be assumed that they are representative of other general sources of charge--flavour violation, so BSM sensitivity is not reduced substantially by excluding this region from \srmet\ or \srjet.
As shown in Figure~\ref{fig:regions}, the region \(\SUMMT<200~\)\GeV\ is designated as \crratio.
If the further requirement of at least one jet with \(\pt>20~\)\GeV\ is placed, the region is designated \crjet, used to study the SM behaviour in a region more kinematically similar to \srjet.
 
The signal-optimised regions make use of three more flavour-symmetric event variables:
\(\METSIG\), \(\MTTWO\) and \(\SUMPT\).
\begin{itemize}
\item
\(\METSIG\) is the so-called `object-based \(\vecptmiss\) significance' defined in Eq.~(15) of Ref.~\cite{ATLAS-CONF-2018-038}. It is a dimensionless measure of the degree to which the apparent missing transverse momentum in the event is `real' (i.e.~attributable to momentum carried away by invisible particles) rather than due to object mismeasurement or pile-up.
\item
\(\MTTWO\equiv\min\limits_{\vec a+\vec b=\vecptmiss}\max\left[m_\mathrm{T}(e, \vec a), m_\mathrm{T}(\mu,\vec b)\right]\) was proposed in Ref.~\cite{Lester:1999tx}, where \(\vec a\) and \(\vec b\) represent the contributions to \ptmiss\ from each semi-leptonic decay of a pair-produced particle, and all possible values that sum to the observed \ptmiss\ are minimised over. It is evaluated using the algorithm of Ref.~\cite{Lester:2014yga}.
\item
\(\SUMPT\equiv |\vec p_\mathrm{T}^e|+|\vec p_\mathrm{T}^\mu|+|\vec p_\mathrm{T}^{j_1}|\) is a simple sum of the magnitudes of the transverse momenta of the two leptons and the most energetic jet in the event.
\end{itemize}
\srrpv\ is defined to require \(\METSIG>10\) and \(\MTTWO>100~\GeV\).
The first requirement anticipates that neutralinos (\(\tilde{\chi}_{1}^{0}\)) of the supersymmetric signals should carry away missing transverse momentum, while the second suppresses SM \(W^+W^-\) backgrounds.
In all other respects, \srrpv\ is identical to \srmet.
 
In contrast, the targeted leptoquark model processes have no invisible particles in the final state, so \srlq\ requires \(\METSIG<6\).
Furthermore, SM backgrounds in this region are suppressed by requiring events to have \(\SUMPT>1~\TeV\). In all other respects, \srlq\ is the same as \srjet.
 
Once the analysis regions are defined, potential biases to the measurement must be considered.
The largest source of strictly one-sided charge--flavour bias in the ratio measurement is the misreconstruction of jets as light leptons in \(W\)+jet events.
In particular: (i) more \(W^+\) than \(W^-\) are produced in LHC proton collisions, and (ii) jets misreconstructed as `fake' leptons are more likely to appear to be electrons than muons.
If uncorrected, these two factors would cause
\(e^-_\text{fake}\mu^+_\text{real}\)
to be more prevalent than
\(e^+_\text{fake}\mu^-_\text{real}\)
and therefore the so-called `fake' background would favour \(\ourRatio<1\).
To remove the bias, a data-driven estimate of the number of fake-lepton events passing any particular selection is determined, separately for each charge combination, and is subtracted from the raw data counts before the ratio \(\ourRatio\) of \(e^+\mu^-\) to \(e^-\mu^+\) counts is calculated.
 
The fake-lepton estimate itself is determined using a Likelihood Matrix Method approach of the form described in Ref.~\cite{Varnes:2016nrb} or `Method B' of Ref.~\cite{Gillam:2014xua}.
This method predicts the fake lepton background where either one or both leptons is fake, and relies on two lepton definitions with different stringencies.
The tighter selection corresponds to the \polished\ definition used in the rest of the analysis, and the looser `\loose' definition relaxes this by removing the isolation requirements, vertex requirements, and loosening the electron identification requirement to the \texttt{Loose} likelihood-based working point defined in Ref.~\cite{EGAM-2018-01}.
Real-lepton efficiencies are derived in \(e^{\pm}e^{\mp}\) and \(\mu^{\pm}\mu^{\mp}\) regions, dominated by \(Z\rightarrow\ell\ell\) events, as the number of events with two \polished\ leptons divided by the number of events with a given lepton loosened to pass the \loose\ requirements.
The fake-lepton efficiencies are derived using a muon tag-and-probe method, using \(\mu^{\pm}\mu^{\pm}\) pairs for the muon efficiency and \(e^{\pm}\mu^{\pm}\) pairs for the electron efficiency.
The `tag' lepton must pass the \polished\ requirements as well as $\pt>50$\,\GeV, and the efficiency is defined as the fraction of \loose\ probe leptons also passing the \polished\ requirements, once the small SM real-lepton background (estimated from MC) has been subtracted.
The requirement for same-charge-sign rather than opposite-charge-sign increases the contribution from events with fake leptons relative to real SM processes, dominantly \(W+\)jets events.
The efficiencies are calculated separately for each lepton flavour and charge (such that the flavour and charge match the lepton that has its selection loosened), and are binned in lepton \pt.
These efficiencies, together with event counts in regions orthogonal to the signal regions and where one lepton is required to pass the \loose\ selection, are used to calculate a prediction for the yield of fake-lepton events in the signal regions. 
The fake-lepton estimate accounts for \(\mathcal{O}(2\%)\) of the events in the signal regions used for the ratio measurement, and \(\mathcal{O}(6\%)\) or \(\mathcal{O}(17\%)\) of the events in the signal region used for the RPV supersymmetry or leptoquark interpretation, respectively.
As can be seen in Figure~\ref{fig:data_both}, the fake-lepton estimate in \emup\ events is indeed generally higher than in \emum\ events.
 
The uncertainty in the fake-lepton estimate includes two uncertainties: one propagated from uncertainty in the values of the efficiencies, and a `non-closure' uncertainty.
The non-closure uncertainty is derived in the region used to calculate electron fake efficiencies: an \(e^{\pm}\mu^{\pm}\) region (with the electron failing the \polished\ selection but passing the \loose\ selection, and the muon passing the `tag' selection described above), which -- like the signal regions -- has fake leptons originating mostly from \(W+\)jet events.
The region is split into two bins, with either \(\SUMMT<200~\)\GeV\ or \(\SUMMT>200~\)\GeV.
The non-closure uncertainty is taken as the fractional difference in event counts in these bins between the total background estimate and the data.
Here, the total background estimate includes the Likelihood Matrix Method fake-background estimate, as well as the real-lepton background estimated using SM MC.
It was checked that the real-lepton background contamination in this region has no significant impact on the uncertainty value.
The non-closure uncertainty has a magnitude of 21\% (13\%) for events with \(\SUMMT>200~\)\GeV\ (\(\SUMMT<200~\)\GeV), which is applied to the signal (control) region.
 
Only two other sources of potential charge--flavour bias motivate application of an explicit correction to data. Firstly, in certain regions of the detector
there are small differences between the reconstruction (and trigger) efficiencies for positively and negatively charged muons.
These are largely a result of the toroidal magnetic field that the muons move in while traversing the muon spectrometer, increasing the relative acceptance of muons of one charge in certain regions, usually anti-symmetrically in \(\eta\).
To remove these differences, weights (depending on muon charge, transverse momentum and pseudorapidity, and derived from \(Z\rightarrow \mu\mu\) samples following the tag-and-probe approach described in Ref.~\cite{MuonPerfFullRun2}) are applied to events after data-taking but before any other use in the analysis.
These weights correct for the bias by taking the efficiency values back to the charge-averaged values.
Approximately two thirds of these weights have values within 1\% of unity.
In addition to introducing an overall acceptance change of \({\sim}0.05\%\),
these weights are responsible for event yields acquiring non-integer values.
Uncertainties associated with this correction are obtained by propagating the statistical uncertainty of the charge-bias measurement.
Secondly, a small correction is applied to data to account for the muon sagitta bias, which is derived in accord with Ref.~\cite{fixedIDTR-2019-05}, and comes with associated uncertainties which are also applied to data.
This charge-dependent bias in muon momentum is caused by misalignment of the detector, and is found to be very minor: 68\% of muons have a bias of less than 0.2\% of the muon \pt.
 
Before unblinding the signal regions, the hypothesis that the proton--proton initial state and experimental effects lead to a bias favouring \(\ourRatio\leq1\) in the SM was confirmed by measuring \(\ourRatio\), binned in `stransverse mass' \MTTWO~\cite{Lester:1999tx}, in \crratio.
Whilst the ratio is consistent with one within uncertainties, the maximal deviation from one is used to define a 2\% `residual-bias' uncertainty encompassing small remaining uncorrected detector biases.
The extrapolation of the uncertainty to high \SUMMT\ was validated by inspecting its impact on the \(\ourRatio\) measurement in \crratio\ and \crjet\ when binned in \SUMMT.
\section{Results}
\label{sec:result}
The observed data and fake-lepton background estimate in the \emum\ and \emup\ channels of \srmet\ and \srjet\ are shown in bins of \MTTWO\ and \SUMPT\ respectively in Figure~\ref{fig:data_both}. Benchmark RPV-supersymmetry or leptoquark signal yields are included to demonstrate that these BSM models favour \emum\ over \emup. In the lower panels of Figure~\ref{fig:data_both}, an estimate of the proportion of SM background processes in each bin is given, showing that \ttbar\ is expected to dominate in most bins apart from the tails, where the fake-lepton, diboson, and single-top backgrounds become proportionally more important.
 
\begin{figure}[htb!]
\centering
\begin{subfigure}[b]{0.49\textwidth}
\includegraphics[width=\textwidth]{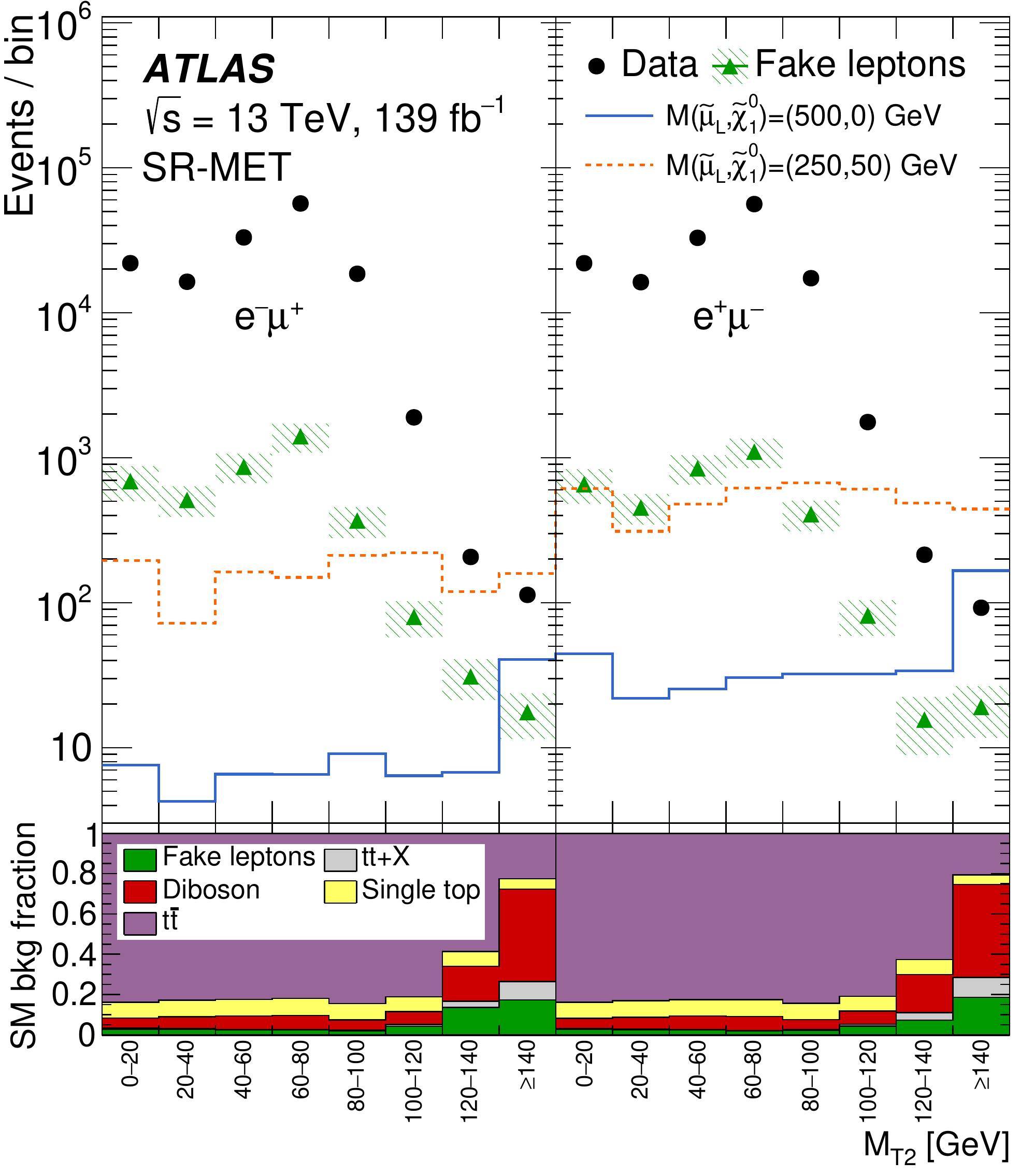}
\caption{\srmet}
\end{subfigure}
\hfill
\begin{subfigure}[b]{0.49\textwidth}
\includegraphics[width=\textwidth]{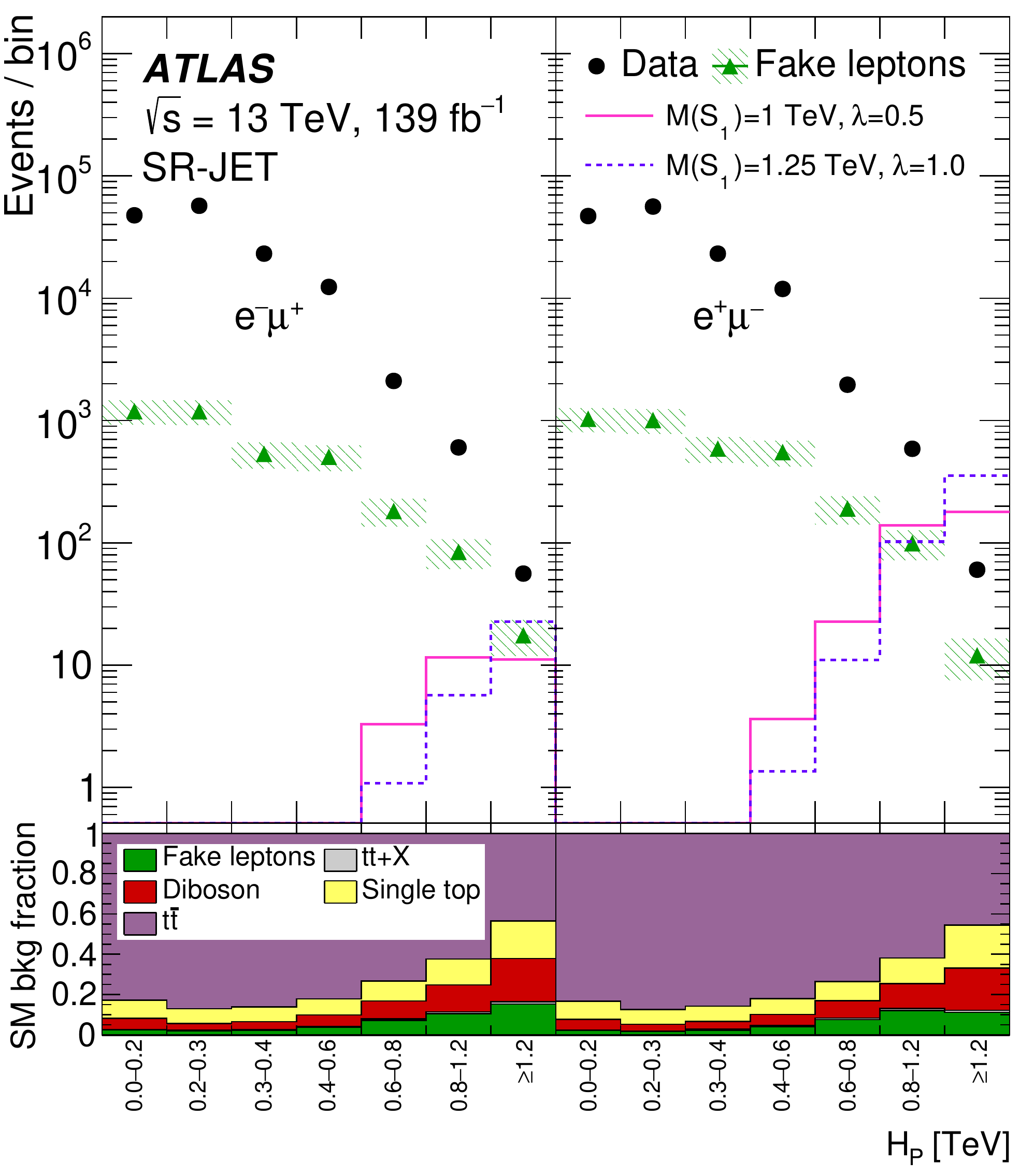}
\caption{\srjet}
\end{subfigure}
\caption{
Distributions of data in the \emum\ and \emup\ channels of the model-independent signal regions, binned in \MTTWO\ for \srmet, and \SUMPT\ for \srjet, to correspond to the ratio measurement binning.
The data has the muon charge and sagitta-bias corrections applied and corresponding uncertainties are added in quadrature in the error bar with the statistical uncertainty of the data.
The fake-lepton background estimate is also shown, along with its uncertainty components added in quadrature, illustrating larger yields in the \emup\ channel as expected.
The lower panel shows the fraction that each SM process contributes to the total SM background in each bin, estimated using standard MC simulations.
The dominant background is \ttbar, whilst the importance of the fake-lepton background increases in higher bins of each variable.
Benchmark RPV-supersymmetry signal models are shown for \srmet, and benchmark scalar leptoquark models are shown for \srjet, which all strongly favour the \emum\ final state over \emup, as expected.
}
\label{fig:data_both}
\end{figure}
 
The ratio, \(\rho\), is measured in bins, \(i\), of \MTTWO~(\(\SUMPT\)) in the \srmet~(\srjet) by maximising a parameterised likelihood model of the observed yields, \(\vec{N}_{\text{obs,i}}^{+-/-+}\). The likelihood model assumes an independent Poisson distribution for the yield in each bin of the charge--flavour channels (\emum\ or \emup):
\begin{align}
\ifOneOrTwoCol{}{&}
\mathcal{L}(\vec{N}_{\text{obs}}^{+-/-+}|\vec{\theta},\vec{\alpha},\vec{\ourRatio})=
\ifOneOrTwoCol{  & }{ \\   }
\ifOneOrTwoCol{   }{ &   }
\prod_{i \in \text{bins}} \left[ \text{Pois}(N_{\text{obs},i}^{-+} \vert w^{-+}_i(\vec{\theta})N_{\text{exp},i} + F_{i}^{-+}(\vec{\alpha})) \right. \nonumber\\
&\left. \times~\text{Pois}(N_{\text{obs},i}^{+-} \vert \ourRatio_i w^{+-}_i(\vec{\theta})N_{\text{exp},i} + F_{i}^{+-}(\vec{\alpha})) \right] \nonumber \\
&\times \prod_{k \in \text{ fake lepton uncertainties}}\text{Gaus}(0|\alpha_{k},1) \nonumber \\
&\times \prod_{j \in \text{ data uncertainties}}\text{Gaus}(0|\theta_{j},1). \numberthis \label{eq:likelihood_rho}
\end{align}
where the expectation in each bin is the combination of a fake-lepton background estimate, \(F_{i}^{+-/-+}\), and a total irreducible (real-lepton) SM expectation, \(N_{\text{exp},i}\), which is a floating parameter in the likelihood.
Uncertainties associated with the Likelihood Matrix Method estimate and the non-closure uncertainty in the fake-lepton background estimate are included by parameterising \(F_{i}^{+-/-+}\) with Gaussian-constrained nuisance parameters, \(\alpha\).
Muon charge and sagitta-bias corrections are already applied to the observed yields,\footnote{These corrections are sufficiently close to unity to support the Poisson modelling assumption for the corrected yields (further evidence can be seen in Table~\ref{tab:postfit:rpvandlq}).} \(\vec{N}_{\text{obs,i}}^{+-/-+}\), with the relative uncertainties on these corrections included in the \(w^{+-/-+}_i(\vec{\theta})\) term and corresponding Gaussian-constrained nuisance parameters on the expected yields, \(\theta_j\).
The `residual-bias' uncertainty is included in the same manner.
A global ratio measurement from combining all bins in a region gives \(\ourRatio=0.987^{+0.022}_{-0.021}\) for \srmet\ and \srjet.
The binned measured ratios (maximum-likelihood estimators of \(\ourRatio_i\)) are shown in Figure~\ref{fig:ratio_with_DATA}.
In the lower bins of these variables the residual-bias uncertainty dominates; in the final two bins of \MTTWO\ and three bins of \SUMPT\ the fake-lepton and statistical uncertainties dominate.
Figure~\ref{fig:ratio_with_DATA} also shows \emph{one-sided} p-values for a hypothesis test of \(\ourRatio=1\) using a modified profile-likelihood-ratio test statistic that equals zero when \(\ourRatio_i\leq1\), calculated using asymptotic approximations~\cite{Cowan:2010js}.
No significant upward deviation from one is seen in any bin, meaning that the SM hypothesis of \(\ourRatio\leq1\) is not excluded anywhere.
The largest upward deviation of \(\ourRatio\) from one has a local significance of \(1\sigma\).
The largest downward deviation from one is \(\ourRatio=0.929^{+0.023}_{-0.022}\), in the \(80-100~\)\GeV\ bin of \MTTWO, with a local significance of \(3.1\sigma\).
The goodness-of-fit significance to the model that \(\ourRatio=1\) in all bins is \(1.6\sigma\), estimated using a likelihood ratio test statistic with the asymptotic approximation.
 
\begin{figure}
\centering
\includegraphics[width=\ifOneOrTwoCol{0.77\linewidth}{0.99\linewidth}]{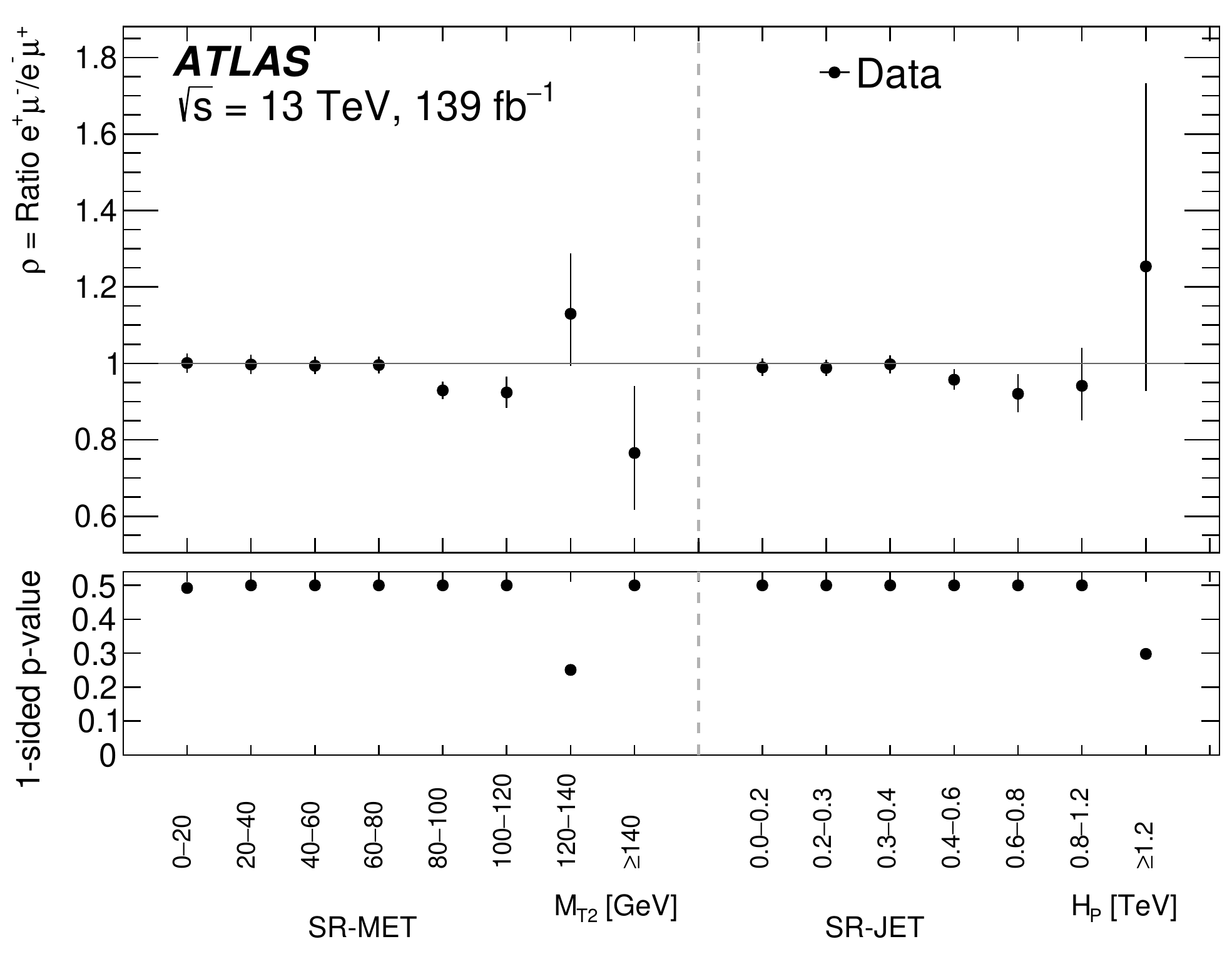}
\caption{A summary of the ratio \(\ourRatio\) measurement in the full Run-2 data for \srmet\ binned in \MTTWO, and \srjet\ binned in \SUMPT.
Muon charge and sagitta-bias corrections are applied to data along with corresponding uncertainties, and the likelihood matrix method is used to estimate the charge--flavour-biased fake-lepton background such that it can be subtracted from the data.
A 2\% uncertainty in \(\ourRatio\), encompassing remaining observed detector biases, is also included.
The lower panel shows the \(p\)-value for a one-sided discovery test to reject the SM hypothesis that \(\ourRatio\leq1\).
}
\label{fig:ratio_with_DATA}
\end{figure}
 
The CL$_\text{s}$ method~\cite{Read:2002hq} is used to obtain 95\% confidence level (CL) upper limits on the number of possible signal events \(S\) entering \srmet\ and \srjet, with a fraction \(z\) entering the \emum\ channel, and these are shown in Figure~\ref{fig:upper_limits} for a range of \(z\) values.
These limits are calculated using a profile-likelihood-ratio test statistic with the likelihood function from Eq.~(\ref{eq:likelihood_rho}) after fixing the ratio values to \(\ourRatio_i=1\) and adding signal components to the Poisson expectations: \(Sz\) in the \emum\ channel, and \(S(1-z)\) in the \emup\ channel, where $S$ is the parameter of interest.
 
\begin{figure}
\centering
\includegraphics[width=\ifOneOrTwoCol{0.77\linewidth}{0.99\linewidth}]{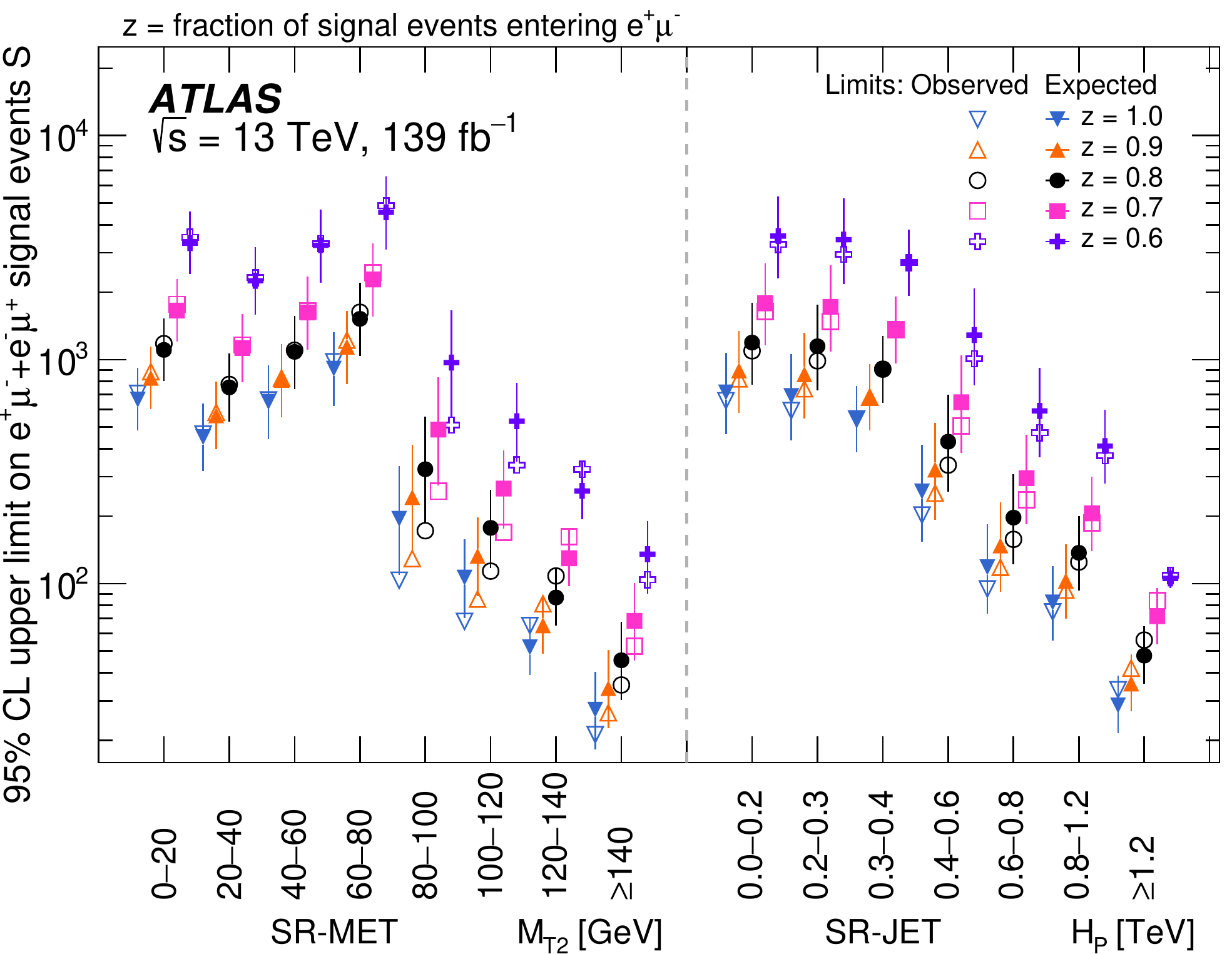}
\caption{Observed and expected 95\% CL upper limits on the total number of signal events entering the \emum\ and \emup\ channels in each bin of \srmet\ and \srjet. The regions are binned in the same way as the ratio (\(\ourRatio\)) measurement.
The limits are shown for a selection of `\(z\)' values, where `\(z\)' is the fraction of the total number of signal events entering the \emum\ channel.
The expected limit is calculated with the asymptotic approximation, by considering when there is a \(50\%\) chance of exclusion at \(95\%\) CL$_\text{s}$, under the hypothesis that \(S=0\) and \(\ourRatio_i=1\).
}
\label{fig:upper_limits}
\end{figure}
 
Having seen consistency with the SM hypothesis in the ratio measurement, limits are placed on parameters of the two benchmark BSM models.
RPV-supersymmetry model exclusion limits are calculated with a one-sided profile-likelihood-ratio test statistic, where a likelihood function is defined for the observed yields in the \emum\ and \emup\ channels of both \srrpv\ and a corresponding control region \crrpv.
This likelihood function is similar to Eq.~(\ref{eq:likelihood_rho}) with signal yield terms added to the Poisson expectations (these signal terms are scaled by a signal strength parameter \(\mu_{\text{sig}}\)) and \(i\) labelling the regions instead of the bins (each region in this model has a single bin).
The \(\ourRatio_i\) are also replaced by a single \(\ourRatio\) that is common to both regions.
In this respect, the control region drives the measurement of \(\ourRatio\) and the signal region determines the value of \(\mu_{\text{sig}}\), which is used as the parameter of interest in the test statistic.
The variance of \(\ourRatio\) across the $(\METSIG,\MTTWO)$-plane (`RPV-plane') outside of the signal region is estimated with measurements of the ratio in bins of the plane axis observables: the binning was chosen to approximately match the statistical precision of the signal region.
The estimated variance of 6\% is added to the likelihood model as an uncertainty in \(\ourRatio\) to cover the modelling assumption that \(\ourRatio\) is invariant across the plane.
This uncertainty covers both statistical and systematic sources of variance.
The model is validated by finding good agreement between the observed and post-fit expected yields in the \emum\ channel of validation regions orthogonal to \srrpv\ and \crrpv\ (defined in Figure~\ref{fig:regions}), where the \emup\ channel of these regions is included in the fit along with both channels of \crrpv.
These validation regions are not included for the test statistic calculations when calculating the limits.
Figure~\ref{fig:plane:rpv} shows the observed and expected yields after the fit in the RPV-plane, demonstrating good agreement in the validation regions.
This analysis is repeated in the $(\SUMPT,\METSIG)$-plane (`LQ-plane') for leptoquark benchmark models (see Figure~\ref{fig:regions}), with an estimated 9\% variance of \(\ourRatio\) across the plane.
Figure~\ref{fig:plane:lq} shows the fit result in the LQ-plane, demonstrating again good agreement in the validation regions.
Uncertainties are included in the signal terms for lepton reconstruction efficiency, energy scale and resolution, and trigger efficiency differences between MC simulation and data; uncertainties in the jet-energy scale and resolution~\cite{Aad:2020flx}, the modelling of the \(\vecptmiss\) soft term~\cite{ATLAS-CONF-2018-023}, and electron charge identification~\cite{Aad:2019tso} are also included.
The RPV signal model yields include theoretical uncertainties in the signal acceptance due to the choice of parton shower model and factorisation and renormalisation scales. The LQ signal models include effects of factorisation and renormalisation scale uncertainties on the NLO cross-section prediction, which form the \(\pm1\sigma_{\text{theory}}\) band on the observed limit.
 
\begin{figure}[htb!]
\centering
\begin{subfigure}[b]{0.49\textwidth}
\includegraphics[width=\textwidth]{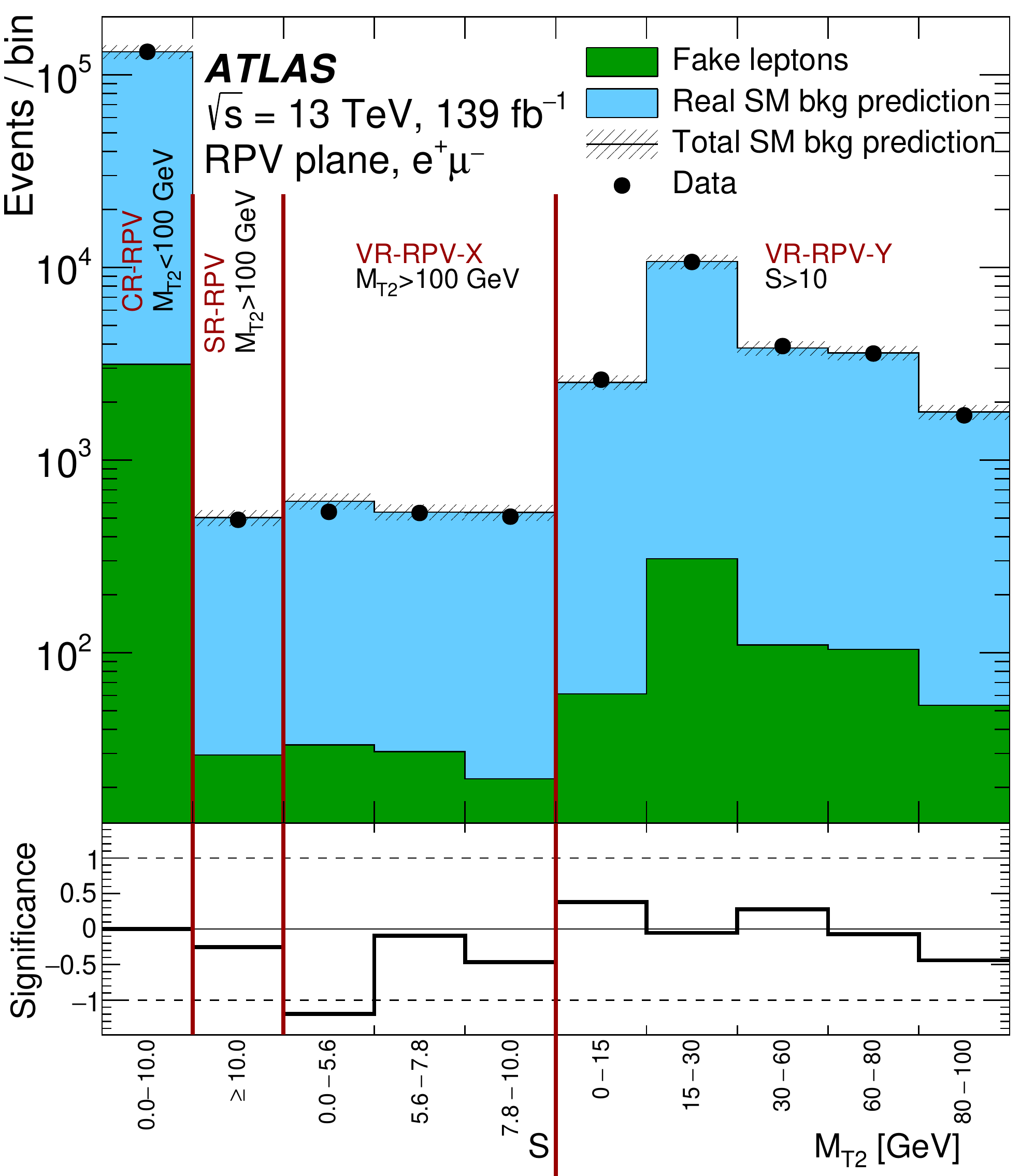}
\caption{RPV-plane}
\label{fig:plane:rpv}
\end{subfigure}
\hfill
\begin{subfigure}[b]{0.49\textwidth}
\includegraphics[width=\textwidth]{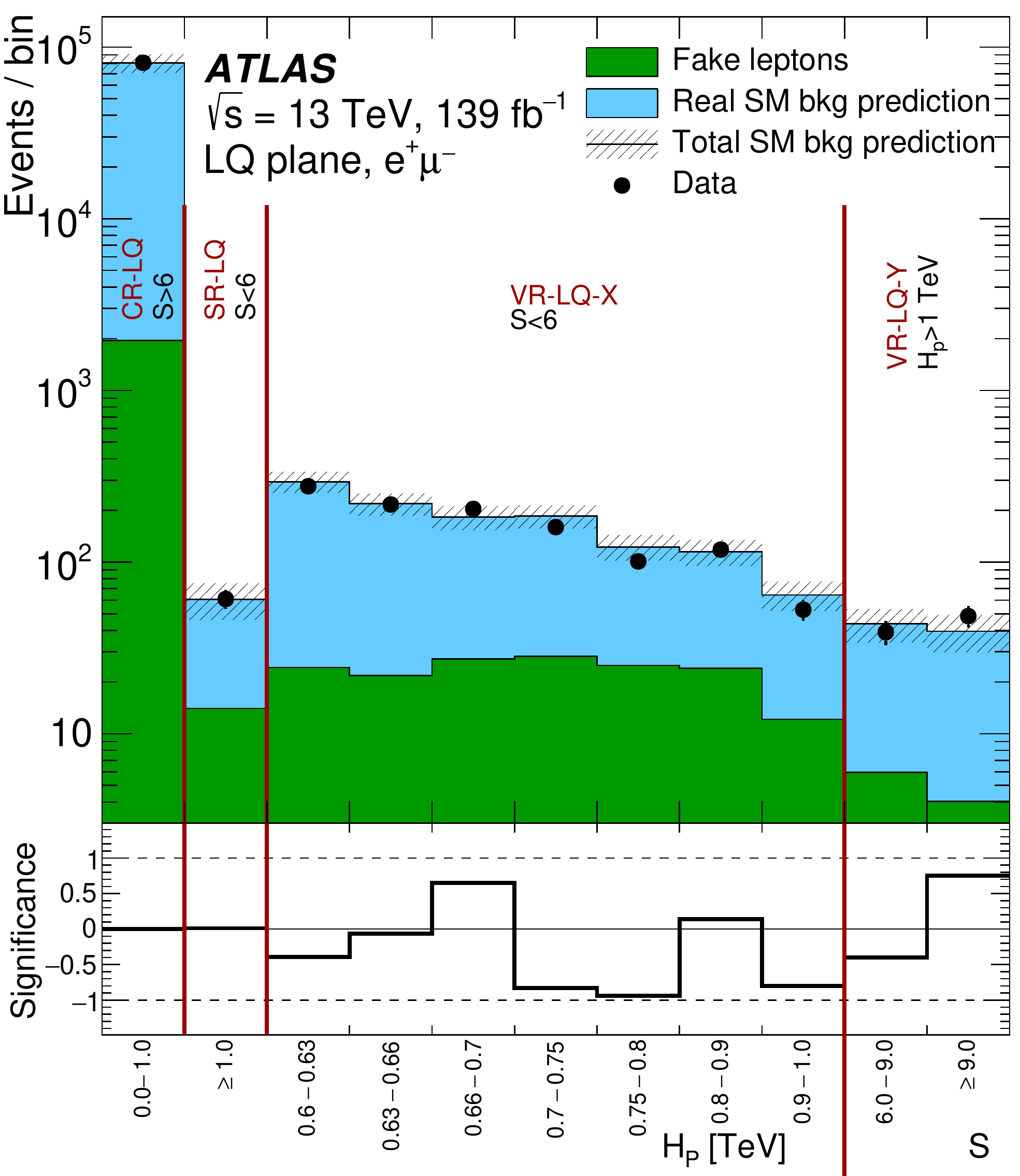}
\caption{leptoquark plane}
\label{fig:plane:lq}
\end{subfigure}
\caption{
Distributions of data in the \emum\ channel of the signal, control, and validation regions, used for the \(R\)-parity-violating supersymmetry model and leptoquark model limit fits.
Note that `VR-LQ-x' has a lower bound of \(\SUMPT>600~\)\GeV, as it is not necessary to validate the ratio extrapolation any further away, kinematically, from the signal region.
The data has the muon charge and sagitta-bias corrections applied.
The fake-lepton background estimate and data-driven real SM background prediction derived from the fit are also shown.
The sum of these is the total SM background estimate, and its error bar includes contributions from the fake-lepton background estimate, muon charge-bias and sagitta-bias uncertainties on the denominator data yields used to fit the SM background estimate.
The lower panel shows the significance of any deviation between the observed data yields and total SM background prediction in each bin, defined in Ref.~\cite{ATL-PHYS-PUB-2020-025}.
}
\end{figure}
 
As shown in Table~\ref{tab:postfit:rpvandlq}, no statistically significant deviations of the data from the total SM background prediction are seen in the \emum\ channels of either signal region.
By construction, since \(N_{\text{exp,i}}\) is a freely floating parameter, good agreement is seen between the data and SM prediction in the \emup\ channels after the fit excluding the \emum\ channels.
These are included in the Table~\ref{tab:postfit:rpvandlq} for completeness and comparison with the benchmark signal yields.
 
\begin{table}[h!]
\centering
\caption{Observed yields, and (post-fit) expected yields for the data-driven SM estimates in the case where \(\mu_{\text{sig}}=0\) and the fit excludes the \emum\ signal region; in such a fit the post-fit expected yields in the \emup\ channel are constrained to match the data exactly.
Additionally, signal yields are shown for the benchmark RPV-supersymmetry signal points in \srrpv\ and the leptoquark signal points in \srlq\ obtained from a fit excluding the \emum\ signal region and setting \(\mu_{\text{sig}}=1\).
Small weights correcting for muon charge biases affect all rows except that containing the fake-lepton estimate.
These weights, \(w_i\), cause non-integer yields.
The uncertainties, \(\sqrt{\sum_i w_i^2}\), are given for data to support the choice made to model the yields with a Poisson distribution.
}
\label{tab:postfit:rpvandlq}
\vspace{0.2cm}
\sisetup{round-mode=figures , round-precision =2, group-digits=integer , group-minimum-digits =5}
\begin{tabular}{l 
S[table-format=4.1, table-alignment=right, round-mode=figures, round-precision=3]
@{$\,\pm\,$}
S[table-format=3.1, table-alignment=right, round-mode=figures, round-precision=2]
@{$\,$} 
S[table-format=4.1, table-alignment=right, round-mode=figures, round-precision=3]
@{$\,\pm\,$} 
S[table-format=2.1, table-alignment=right, round-mode=figures, round-precision=2]
@{$\,$}
S[table-format=4.1, table-alignment=right, round-mode=figures, round-precision=3]
@{$\,\pm\,$} 
S[table-format=2.1, table-alignment=right, round-mode=figures, round-precision=2]
@{$\,$}
S[table-format=4.1, table-alignment=right, round-mode=figures, round-precision=3]
@{$\,\pm\,$} 
S[table-format=2.1, table-alignment=right, round-mode=figures, round-precision=2]
@{$\,$}
}
\toprule
& \multicolumn{4}{c}{\srrpv}
& \multicolumn{4}{c}{\srlq} \\
& \multicolumn{2}{c}{\(e^+ \mu^-\)} & \multicolumn{2}{c}{\(e^- \mu^+\)}
& \multicolumn{2}{c}{\(e^+ \mu^-\)} & \multicolumn{2}{c}{\(e^- \mu^+\)}
\\
\hline \hline
\(M(\tilde{\chi}_{1}^{0},\tilde{\mu})=(0,500)\)~\GeV, \(\lptwothreeone=1\) & 191 & 23 & 46.8 & 7.7 \\ 
\(M(\tilde{\chi}_{1}^{0},\tilde{\mu})=(50,250)\)~\GeV, \(\lptwothreeone=1\) & 1160 & 130 & 361 & 97 \\ 
\(M(S_{1})\) = 1 \TeV, \(\lambda\) = 0.5 & \multicolumn{4}{c}{} & 214 & 15  & 14.5 & 1.8 \\ 
\(M(S_{1})\) = 1.25 \TeV, \(\lambda\) = 1.0 & \multicolumn{4}{c}{} & 356 & 53  & 22.9 & 3.7 \\ 
\hline
Data
& 489 & 22 & 510 & 23
& 60.9 & 7.8 & 69.1 & 8.3 \\
\hline
Total SM expectation & 503 & 48 & 510 & 26
& 61 & 15 & 69 & 12 \\
\hphantom{m} $\bullet$ part due to real leptons
& 473 & 47 & 479 & 24
& 47 & 13 & 47 & 11  \\
\hphantom{m} $\bullet$ part due to fake leptons
& 29.4 & 8.2 & 30.3 & 8.3
& 14.1 & 6.5 & 22.1 & 6.6 \\
\bottomrule
\end{tabular}
 
\end{table}
 
\begin{figure}[htb!]
\centering
\includegraphics[width=\ifOneOrTwoCol{0.77\linewidth}{0.99\linewidth}]{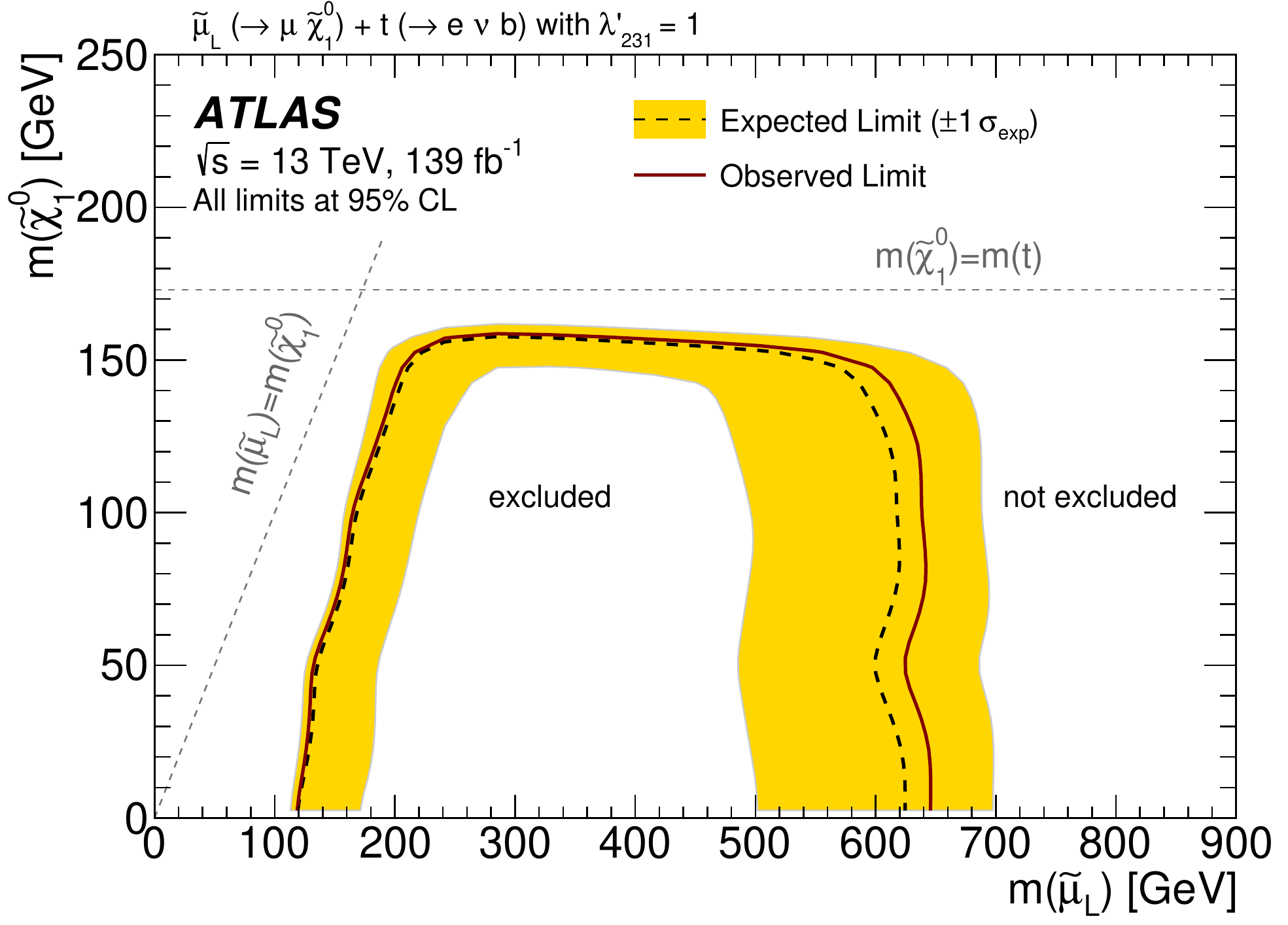}
\caption{
\label{subfig:reach_RPV}
Expected and observed exclusion limits are shown for RPV-supersymmetry models which allow for production of a single smuon (decaying into a muon and neutralino) in association with a top quark (decaying leptonically).
The expected limit is calculated with the asymptotic approximation, by considering when there is a \(50\%\) chance of exclusion at \(95\%\) CL$_\text{s}$, under the SM-only hypothesis.
The smuon is produced through the \lptwothreeone\ coupling, which is fixed at unity.
All limits are computed at 95\% CL and all uncertainties are included.
Also shown are dotted lines to indicate the two kinematic limits for the RPV process considered.
}
\end{figure}
 
\begin{figure}[htb!]
\centering
\includegraphics[width=\ifOneOrTwoCol{0.77\linewidth}{0.99\linewidth}]{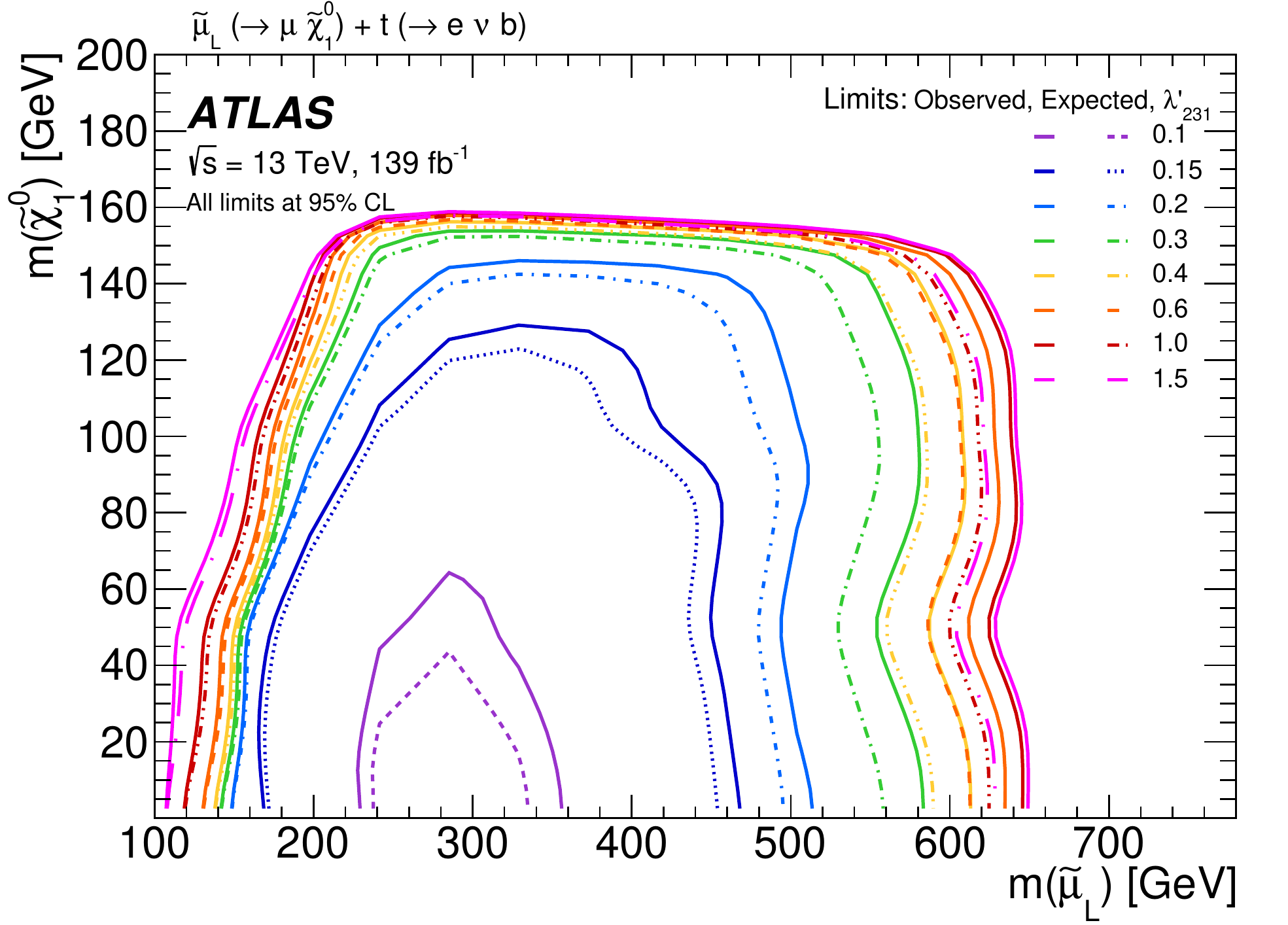}
\caption{
\label{subfig:reach_RPV_coupling}
Expected and observed exclusion limits are shown for RPV-supersymmetry models which allow for production of a single smuon (decaying into a muon and neutralino) in association with a top quark (decaying leptonically).
The expected limits is calculated with the asymptotic approximation, by considering when there is a \(50\%\) chance of exclusion at \(95\%\) CL$_\text{s}$, under the SM-only hypothesis.
The smuon is produced through the \lptwothreeone\ coupling, which takes values up to 1.5. All limits are computed at 95\% CL and all uncertainties are included.
}
\end{figure}
 
The observed and expected RPV-supersymmetry limits are shown in Figure~\ref{subfig:reach_RPV} for the case where the \lptwothreeone\ coupling is fixed at one, and in Figure~\ref{subfig:reach_RPV_coupling} where the coupling takes values between 0.1 and 1.5.
The perturbative upper limit for the \lptwothreeone\ coupling is 1.12 at the \(Z\)-boson mass~\cite{Allanach:1999ic}, and increases with the energy scale.
For coupling values above 1, the limit at high smuon mass becomes constant since the cross-section increase and the branching-ratio decrease cancel each other out.
Neutralino masses near and above the top-quark mass are not excluded, as here the neutralino can decay through the RPV coupling and no real \(\vecptmiss\) remains in the final state.
For the largest coupling value considered, smuon masses up to \(650~\GeV\) are excluded.
 
\begin{figure}[htb!]
\centering
\includegraphics[width=\ifOneOrTwoCol{0.74\linewidth}{0.99\linewidth}]{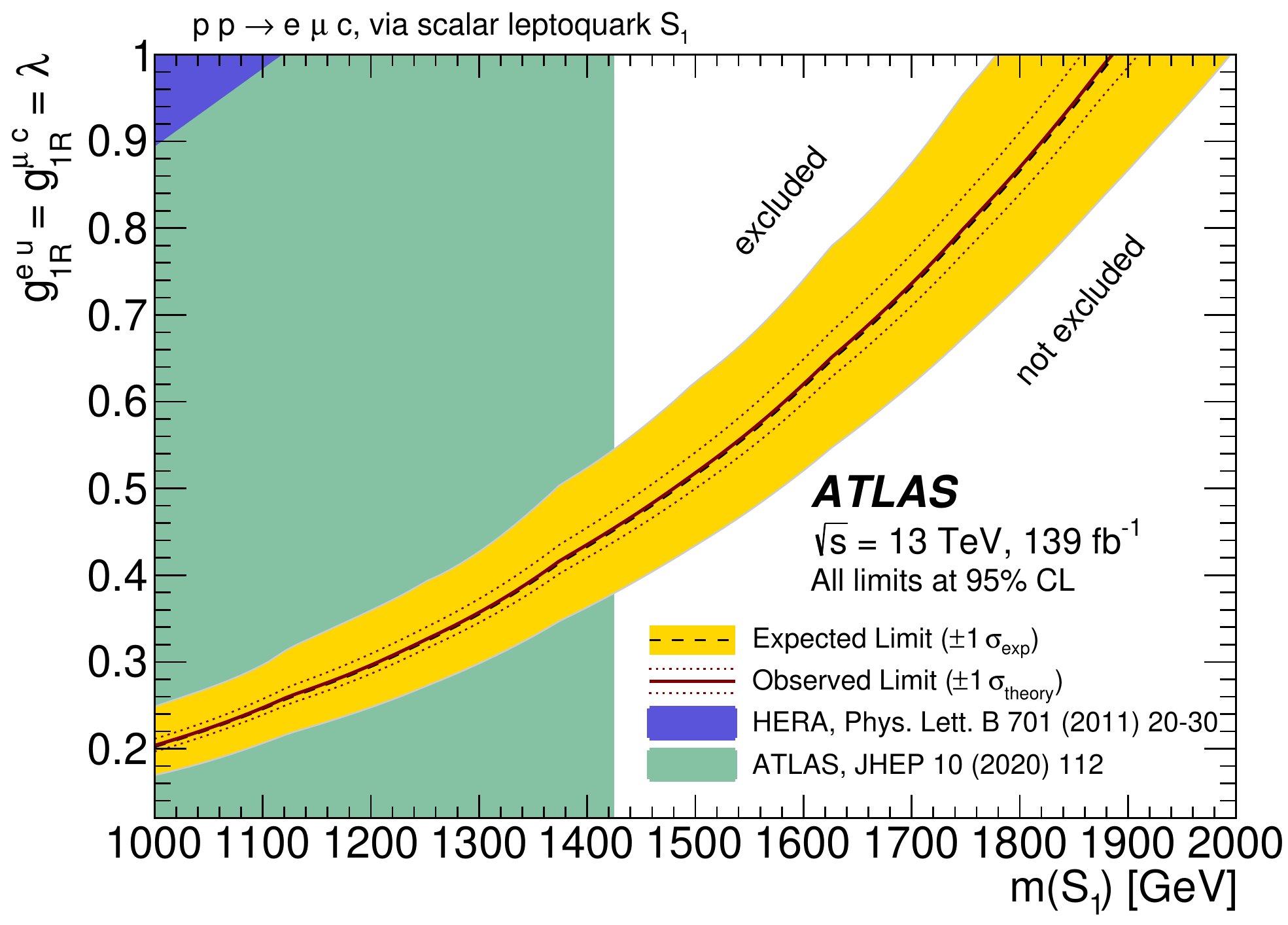}
\caption{
\label{subfig:reach_LQ}
Expected and observed exclusion limits are shown for models featuring production of a single scalar leptoquark which decays into a muon and charm quark.
All limits are computed at 95\% CL and all uncertainties are included.
The \(\pm1 \sigma_{\text{theory}}\) band considers the theoretical uncertainty of the NLO cross section used for the leptoquark signals.
The expected limit is calculated with the asymptotic approximation, by considering when there is a \(50\%\) chance of exclusion at \(95\%\) CL$_\text{s}$, under the SM-only hypothesis.
The underlaid exclusion is derived from a previous ATLAS leptoquark pair-production search \cite{Aad:2020iuy}, considering the result from $\mu c$ final states.
Since the model in this analysis requires leptoquarks to have two decay modes while that of Ref.~\cite{Aad:2020iuy} assumes only one, a branching ratio of 50\% into $\mu c$ was used when determining the position of the exclusion boundary.
The interpretation of Ref.~\cite{Aad:2020iuy} also assumes that the narrow-width approximation is valid for leptoquarks over the range of coupling values shown.}
\end{figure}
 
Figure~\ref{subfig:reach_LQ} shows the observed and expected limits on the leptoquark models considered.
Since the energy required to produce a pair of leptoquarks is always double that required to make a single one, the suppression of high centre-of-mass energies by steeply falling parton distribution functions naturally leads to places where this analysis has better reach in leptoquark mass than analyses which have targeted pair production.\footnote{This benefit comes, though, at the cost of requiring two non-zero leptoquark couplings \( g_\textrm{1R}^{eu}\ne 0 \ne g_\textrm{1R}^{\mu c} \) rather than just the one assumed by existing pair--production searches.}
Notably, leptoquark couplings of \( g_\textrm{1R}^{eu}=g_\textrm{1R}^{\mu c}>0.46 \) are newly excluded for masses above \(1420~\GeV\), up to a value of unity (the largest coupling considered) for a leptoquark mass of \(1880~\GeV\).
\FloatBarrier
\section{Conclusion}
To search for evidence of new physics, this analysis compares the production cross sections for  \(e^+ \mu^-\) and \(e^- \mu^+\) by investigating the ratio  \(\ourRatio= \frac{\sigma(pp\rightarrow e^+\mu^-+X)}{\sigma(pp\rightarrow e^-\mu^++X)}\) in a variety of signal regions.
New physics processes could potentially raise or lower \(\ourRatio\) from one, but even though the largest Standard Model effect known to lower \(\ourRatio\) was subtracted, the model-independent tests presented in the first half of this analysis look only for evidence of the `unexpected' scenario of \(\ourRatio>1\).
No significant model-independent evidence for \(\ourRatio>1\) was seen when analysing $139~\ifb$ of proton--proton collision data recorded at $\sqrt s = 13~\TeV$ by the ATLAS detector at the LHC.
 
Further observations were conducted in more exclusive regions optimised for particular signals beyond the Standard Model.
These regions targeted: (i) \(R\)-parity-violating supersymmetric models with non-zero \lptwothreeone\ couplings, with smuons and stable neutralinos, and (ii) scalar leptoquark models with \(g_\textrm{1R}^{eu}=g_\textrm{1R}^{\mu c}\).
The secondary measurements were then used to create exclusions in planes in sparticle or leptoquark model spaces.
The search was able to exclude singly produced smuons in certain models in which the only other light sparticle is a neutralino, albeit with those exclusions dependent on the existence of \lptwothreeone\ \(R\)-parity-violating couplings.
Scalar leptoquarks with \(g_\textrm{1R}^{eu}=g_\textrm{1R}^{\mu c}\leq1\) were excluded for masses below \(1880~\GeV\).
This value reduces to \(g_\textrm{1R}^{eu}=g_\textrm{1R}^{\mu c}=0.46\) for \(1420~\GeV\) (close to the limits obtained in analyses based on leptoquark pair production).

\section*{Acknowledgements}


We thank CERN for the very successful operation of the LHC, as well as the
support staff from our institutions without whom ATLAS could not be
operated efficiently.
 
We acknowledge the support of
ANPCyT, Argentina;
YerPhI, Armenia;
ARC, Australia;
BMWFW and FWF, Austria;
ANAS, Azerbaijan;
SSTC, Belarus;
CNPq and FAPESP, Brazil;
NSERC, NRC and CFI, Canada;
CERN;
ANID, Chile;
CAS, MOST and NSFC, China;
Minciencias, Colombia;
MEYS CR, Czech Republic;
DNRF and DNSRC, Denmark;
IN2P3-CNRS and CEA-DRF/IRFU, France;
SRNSFG, Georgia;
BMBF, HGF and MPG, Germany;
GSRI, Greece;
RGC and Hong Kong SAR, China;
ISF and Benoziyo Center, Israel;
INFN, Italy;
MEXT and JSPS, Japan;
CNRST, Morocco;
NWO, Netherlands;
RCN, Norway;
MEiN, Poland;
FCT, Portugal;
MNE/IFA, Romania;
JINR;
MES of Russia and NRC KI, Russian Federation;
MESTD, Serbia;
MSSR, Slovakia;
ARRS and MIZ\v{S}, Slovenia;
DSI/NRF, South Africa;
MICINN, Spain;
SRC and Wallenberg Foundation, Sweden;
SERI, SNSF and Cantons of Bern and Geneva, Switzerland;
MOST, Taiwan;
TAEK, Turkey;
STFC, United Kingdom;
DOE and NSF, United States of America.
In addition, individual groups and members have received support from
BCKDF, CANARIE, Compute Canada and CRC, Canada;
COST, ERC, ERDF, Horizon 2020 and Marie Sk{\l}odowska-Curie Actions, European Union;
Investissements d'Avenir Labex, Investissements d'Avenir Idex and ANR, France;
DFG and AvH Foundation, Germany;
Herakleitos, Thales and Aristeia programmes co-financed by EU-ESF and the Greek NSRF, Greece;
BSF-NSF and GIF, Israel;
Norwegian Financial Mechanism 2014-2021, Norway;
NCN and NAWA, Poland;
La Caixa Banking Foundation, CERCA Programme Generalitat de Catalunya and PROMETEO and GenT Programmes Generalitat Valenciana, Spain;
G\"{o}ran Gustafssons Stiftelse, Sweden;
The Royal Society and Leverhulme Trust, United Kingdom.
 
The crucial computing support from all WLCG partners is acknowledged gratefully, in particular from CERN, the ATLAS Tier-1 facilities at TRIUMF (Canada), NDGF (Denmark, Norway, Sweden), CC-IN2P3 (France), KIT/GridKA (Germany), INFN-CNAF (Italy), NL-T1 (Netherlands), PIC (Spain), ASGC (Taiwan), RAL (UK) and BNL (USA), the Tier-2 facilities worldwide and large non-WLCG resource providers. Major contributors of computing resources are listed in Ref.~\cite{ATL-SOFT-PUB-2021-003}.
 
 
\section*{Appendix \appsmmc}
The top pair and single-top backgrounds were modelled using \textsc{Powheg\,Box}~\CiteRef{Powheg}~v2 interfaced to \PythiaEight\ \CiteRef{PythiaEight} and \EvtGen\ \cite{Lange:2001uf} and the \NNPDFTwoThree\ \CiteRef{NNPDF}\ PDF.
A dilepton filter was applied to the \ttbar\ and \tw\ processes.
 
The diboson backgrounds were modelled using \Sherpa\ \CiteRef{Sherpa}.
Hard processes with no or one additional jet in the final state were simulated at NLO, while up to three additional jets were included at LO.
\Sherpa\,2.2.2 was used for the fully leptonic final states (\llll, \lllv, \llvv\ and \lvvv) together with the \CTTen\ PDF \CiteRef{CTTen}.
For the semileptonic final states (\llqq\ and \lvqq), \Sherpa\,2.2.1 was used with the \NNPDF~\CiteRef{NNPDF}\ PDF.
The loop-induced processes (\ggllll, \ggllvv, \lllljj, \lllvjj\ and same-sign \llvvjj) were generated using \Sherpa\,2.1.1 and the \CTTen\ PDF.
 
The \zjets\ background was modelled using \Sherpa\,2.2.1 with the \NNPDF\ PDF.
Up to two jets were generated at NLO and up to four at LO.
 
The \ttX\ processes were simulated  using \SPLITaMcAtNloPythiaEight.
The \EvtGen\ program was used for properties of the bottom and charm hadron decays.
The events are normalised to their respective NLO cross sections.
\printbibliography

\clearpage
 
\begin{flushleft}
\hypersetup{urlcolor=black}
{\Large The ATLAS Collaboration}

\bigskip

\AtlasOrcid[0000-0002-6665-4934]{G.~Aad}$^\textrm{\scriptsize 98}$,    
\AtlasOrcid[0000-0002-5888-2734]{B.~Abbott}$^\textrm{\scriptsize 124}$,    
\AtlasOrcid[0000-0002-7248-3203]{D.C.~Abbott}$^\textrm{\scriptsize 99}$,    
\AtlasOrcid[0000-0002-2788-3822]{A.~Abed~Abud}$^\textrm{\scriptsize 34}$,    
\AtlasOrcid[0000-0002-1002-1652]{K.~Abeling}$^\textrm{\scriptsize 51}$,    
\AtlasOrcid[0000-0002-2987-4006]{D.K.~Abhayasinghe}$^\textrm{\scriptsize 91}$,    
\AtlasOrcid[0000-0002-8496-9294]{S.H.~Abidi}$^\textrm{\scriptsize 27}$,    
\AtlasOrcid[0000-0002-9987-2292]{A.~Aboulhorma}$^\textrm{\scriptsize 33e}$,    
\AtlasOrcid[0000-0001-5329-6640]{H.~Abramowicz}$^\textrm{\scriptsize 157}$,    
\AtlasOrcid[0000-0002-1599-2896]{H.~Abreu}$^\textrm{\scriptsize 156}$
% ...
\bigskip
\\

% ...
$^{27}$Physics Department, Brookhaven National Laboratory, Upton NY; United States of America.\\
% ...
$^{33}$$^{(a)}$Facult\'e des Sciences Ain Chock, R\'eseau Universitaire de Physique des Hautes Energies - Universit\'e Hassan II, Casablanca;$^{(b)}$Facult\'{e} des Sciences, Universit\'{e} Ibn-Tofail, K\'{e}nitra;$^{(c)}$Facult\'e des Sciences Semlalia, Universit\'e Cadi Ayyad, LPHEA-Marrakech;$^{(d)}$LPMR, Facult\'e des Sciences, Universit\'e Mohamed Premier, Oujda;$^{(e)}$Facult\'e des sciences, Universit\'e Mohammed V, Rabat;$^{(f)}$Mohammed VI Polytechnic University, Ben Guerir; Morocco.\\
$^{34}$CERN, Geneva; Switzerland.\\
% ...
$^{51}$II. Physikalisches Institut, Georg-August-Universit\"{a}t G\"ottingen, G\"ottingen; Germany.\\
% ...
$^{91}$Department of Physics, Royal Holloway University of London, Egham; United Kingdom.\\
% ...
$^{98}$CPPM, Aix-Marseille Universit\'e, CNRS/IN2P3, Marseille; France.\\
$^{99}$Department of Physics, University of Massachusetts, Amherst MA; United States of America.\\
% ...
$^{124}$Homer L. Dodge Department of Physics and Astronomy, University of Oklahoma, Norman OK; United States of America.\\
% ...
$^{156}$Department of Physics, Technion, Israel Institute of Technology, Haifa; Israel.\\
$^{157}$Raymond and Beverly Sackler School of Physics and Astronomy, Tel Aviv University, Tel Aviv; Israel.\\
% ...
$^{a}$ Also at Borough of Manhattan Community College, City University of New York, New York NY; United States of America.\\
$^{b}$ Also at Bruno Kessler Foundation, Trento; Italy.\\
$^{c}$ Also at Center for High Energy Physics, Peking University; China.\\
$^{d}$ Also at Centro Studi e Ricerche Enrico Fermi; Italy.\\
$^{e}$ Also at CERN, Geneva; Switzerland.\\
$^{f}$ Also at D\'epartement de Physique Nucl\'eaire et Corpusculaire, Universit\'e de Gen\`eve, Gen\`eve; Switzerland.\\
% ...

\end{flushleft}


\end{document}